\documentclass[11pt,preprint]{aastex} 

\slugcomment{Accepted to the Astrophysical Journal Supplement Series}

\defcitealias{GPB-I}{Paper~I}
\defcitealias{GPB-II}{Paper~II}
\defcitealias{GPB-III}{Paper~III}
\defcitealias{GPB-IV}{Paper~IV}
\defcitealias{GPB-V}{Paper~V}
\defcitealias{GPB-VI}{Paper~VI}
\defcitealias{GPB-VII}{Paper VII}
\newcommand{\GPB}{\mbox{\em GP-B}}
\newcommand{\IMP}{\mbox{IM~Peg}}
\newcommand{\Ba}{\objectname[ICRF J225307.3+194234]{B2250+194}}
\newcommand{\Bb}{\objectname[87GB 225231.0+171747]{B2252+172}}
\newcommand{\C}{\objectname[3C454.3]{\mbox{3C~454.3}}}

\newcommand{\masyr}{\mbox{mas~yr$^{-1}$}}
\newcommand{\masyryr}{\mbox{mas~yr$^{-2}$}}

\newcommand{\Rsol}{\mbox{R$_\sun$}}

\newcommand{\mua}{\mbox{$\mu_{\alpha*}$}}
\newcommand{\mud}{\mbox{$\mu_{\delta}$}}
\newcommand{\dotmua}{\mbox{$\dot\mu_{\alpha*}$}}
\newcommand{\dotmud}{\mbox{$\dot\mu_{\delta}$}}

\newcommand{\Ra}[4]{\mbox{${#1}^{\rm h} \; {#2}^{\rm m} \; {#3}\fs{#4} $}}
\newcommand{\dec}[4]{\mbox{${#1}\arcdeg \; {#2}\arcmin \; {#3}\farcs{#4} $}}
\newcommand{\RA}{$\alpha$} 
\newcommand{\RhA}{$\alpha$} 
\newcommand{\DEC}{$\delta$} 
\newcommand{\Aca}{\mbox{$A_{\rm c\alpha}$}}
\newcommand{\Acd}{\mbox{$A_{\rm c\delta}$}}
\newcommand{\Asa}{\mbox{$A_{\rm s\alpha}$}}
\newcommand{\Asd}{\mbox{$A_{\rm s\delta}$}}

\newcommand{\pone}{\mbox{Paper~I}}
\newcommand{\ptwo}{\mbox{Paper~II}}
\newcommand{\pthree}{\mbox{Paper~III}}
\newcommand{\pfour}{\mbox{Paper~IV}}

\newcommand{\psix}{\mbox{Paper~VI}}
\newcommand{\pseven}{\mbox{Paper~VII}}

\shorttitle{VLBI for {\em Gravity Probe~B}.  V.}
\shortauthors{Ratner et al.}

\begin{document}
 
\title{VLBI for {\em Gravity Probe B}.\\ 
V. Proper Motion and Parallax of the Guide Star, IM Pegasi}

\author{M. I. Ratner\altaffilmark{1},
N. Bartel\altaffilmark{2}, M. F. Bietenholz\altaffilmark{2,3},
D. E. Lebach\altaffilmark{1},
J.-F. Lestrade\altaffilmark{4}, 
R. R. Ransom\altaffilmark{2,5}, and
I. I. Shapiro\altaffilmark{1}}

\altaffiltext{1}{Harvard-Smithsonian Center for Astrophysics, 60
Garden Street, Cambridge, MA 02138, USA}

\altaffiltext{2}{Department of Physics and Astronomy, York University,
4700 Keele Street, Toronto, ON M3J 1P3, Canada}

\altaffiltext{3}{Now also at Hartebeesthoek Radio Astronomy Observatory,
PO Box 443, Krugersdorp 1740, South Africa}

\altaffiltext{4}{Observatoire de Paris, Centre national de la
recherche scientifique, 77 Av.\ Denfert Rochereau, 75014 Paris,
France}

\altaffiltext{5}{Now at Okanagan College, 583 Duncan Avenue West, 
Penticton, B.C., V2A 2K8, Canada and also at the National Research Council 
of Canada, Herzberg Institute of Astrophysics, Dominion Radio 
Astrophysical Observatory, P.O. Box 248, Penticton, B.C., V2A 6K3, 
Canada}

\begin{abstract} 

We present the principal astrometric results of the very-long-baseline
interferometry (VLBI) program undertaken in support of the {\em
Gravity Probe~B} (\GPB) relativity mission.  VLBI observations of the
\GPB\ guide star, the RS~CVn binary IM~Pegasi (HR~8703), yielded
positions at 35~epochs between 1997 and 2005.  We discuss the
statistical assumptions behind these results and our methods for
estimating the systematic errors.  We find the proper motion of \IMP\
in an extragalactic reference frame closely related to the
International Celestial Reference Frame 2 (ICRF2) to be $-20.83$ $\pm$
0.03 $\pm$ 0.09 \masyr\ in right ascension and $-27.27$ $\pm$ 0.03
$\pm$ 0.09 \masyr\ in declination.  For each component the first
uncertainty is the statistical standard error and the second is the
total standard error (SE) including plausible systematic errors.  We
also obtain a parallax of 10.37 $\pm$ 0.07~mas (distance: $96.4 \pm
0.7$~pc), for which there is no evidence of any significant
contribution of systematic error.  Our parameter estimates for the
$\sim$25-day-period orbital motion of the stellar radio emission have
SEs corresponding to $\sim$0.10~mas on the sky in each coordinate.  
The total SE of our estimate of \IMP's proper motion is  
$\sim$30\% smaller than the accuracy goal set by the \GPB\ 
project before launch:  0.14~\masyr\ for 
each coordinate of \IMP's proper motion.
Our results ensure that the uncertainty in \IMP's proper motion
makes only a very small contribution to the uncertainty of the \GPB\
relativity tests. 

\end{abstract}


\section{Introduction}
\label{sintro}

This paper is the fifth in a series describing the astronomical effort
undertaken in support of the NASA/Stanford {\em Gravity Probe B}
(\GPB) relativity mission, an Earth-orbiting mission to test the
geodetic and frame-dragging predictions of general relativity.  As
described in \pone\ \citep{GPB-I}, the rotating \GPB\ spacecraft
monitored the precessions of four ultrahigh accuracy on-board 
gyroscopes with respect to the spacecraft.  To transform these
precessions to a reference frame that is not rotating with respect to
the distant universe, the mission team required both the star-tracking
data recorded by the spacecraft's telescope and independent knowledge
of the proper motion of an adequately bright ``guide" star.  Before
the launch of \GPB\ , the team set the accuracy requirement for that
star's proper motion at 0.14~milliarcseconds per year (\masyr)
standard error (SE) in each coordinate.  Since the proper motion of no
bright star was known with such accuracy, we undertook a dedicated
program of astrometry to determine this proper motion for the chosen
guide star, IM~Pegasi (HR~8703).  This star is an RS~Canum Venaticorum
(RS~CVn) spectroscopic binary star, with orbital period $\sim$25~d and
variable radio emission at centimeter wavelengths.  The basic
properties of \IMP, and the requirements and process that led to its
selection, are also described in \citetalias{GPB-I}.  To achieve the
required astrometric accuracy, we observed \IMP\ using the radio
astronomical technique of very-long-baseline interferometry (VLBI) at
35 epochs over a span of 8.5 years.

In \S~\ref{sobs}, we briefly describe our VLBI observations, while
referencing, as appropriate, the earlier papers of this series.  Next,
in \S~\ref{spos}, we outline the process by which we estimate, for
each session of observations, a single effective position for the
stellar radio emission.  We also comment on the most important sources
of error in this process.  In \S~\ref{ssol} we describe how we
estimate the astrometric parameters of \objectname[HR 8703] \IMP\ from
this set of radio positions, then present the resulting parameter
estimates and postfit position residuals, and proceed to discuss the
goodness-of-fit of our model and our estimates of the statistical and
systematic errors.  We discuss our final results in \S~\ref{sfinres},
and compare them with those from the {\em Hipparcos} mission as well
as from ground-based optical observations in \S~\ref{scomp}.  In
\S~\ref{sconc}, we summarize our conclusions.  Throughout, we use the
words ``images'' and ``maps" almost interchangeably.

\section{Observations}
\label{sobs}

We designed our VLBI program to meet the requirements of the \GPB\
mission.  One important consideration was that we could not rule out
the possibility that \IMP\ is part of a larger triple or multiple
system, and therefore would have a time-dependent apparent proper
motion.  Consequently, we decided to make enough VLBI observations,
especially during the years immediately before and after the launch of
\GPB, to ensure that the proper-motion requirement could be met were
there a nearly constant rate of ``proper acceleration" (see
\S~\ref{ssmodel}).  Moreover, from the same set of observations we
needed to determine \IMP's parallax and the orbital motion of the
radio emission, expected to be associated mainly with only one of the
stellar components of the spectroscopic binary.  From the time of the
selection of \IMP\ as the guide star in 1997
\citepalias[see][]{GPB-I}, we scheduled, made, and analyzed about four
sessions of VLBI observations every year until the \GPB\ mission
ended.  We thus conducted 35 sessions of VLBI observations of \IMP\
between January 1997 and July 2005.

In each session, observations of \IMP\ were interleaved (every 5.5 to 7.3 minutes)
in a repeated cycle with observations of two or three extragalactic reference
sources (see below).  For most sessions we achieved excellent $u$-$v$
coverage, good angular resolution, and high sensitivity through the use of 
full tracks with all available VLBA stations, the VLA, the Effelsberg 100~m
antenna, and the three 70~m antennas of NASA's Deep Space Network (DSN).  
Our maximum projected baseline length for most sessions was $\sim$8,900~km.
During our first two sessions we recorded only right-circular polarization, with
a bit rate of 112 Mbits s$^{-1}$.  During the remaining sessions 
we recorded both circular polarizations, with a total bit rate of 
128 Mbits s$^{-1}$ in all but the last three of our 35 sessions, for which
a 256 Mbits s$^{-1}$ rate was used.
For a full description of
our array, typical scan lengths, and data recording parameters, see
\ptwo\ \citep{GPB-II}.  The high sensitivity
of this array, together with our use of the technique of phase-referenced
mapping, allowed us to map \IMP\ even when its flux density fell to as low
as 0.2~mJy.  All the results analyzed below were obtained from observations at
$\lambda\ =$ 3.6~cm, for which the synthesized beam size was typically 
$\sim$1~mas $\times \sim$2~mas.  During all but a few sessions, we used only this band, to
maximize our detection sensitivity to the unpredictable and 
highly variable emission from \IMP, while still
cycling rapidly among the sources so as to facilitate phase referencing
and reduce many sources of astrometric error.  Our strongest and closest
reference source, the 7-10~Jy radio-bright quasar \C, lying 0.7\arcdeg\
mostly south and somewhat east of \IMP, was observed in all of our sessions, 
as was the 0.35-0.45~Jy quasar \Ba\ (ICRF J225307.3+194234), which lies 2.9\arcdeg\ north of the star. 
During the final 12 sessions, we also included in our observing cycle the 
0.017~Jy compact source \Bb\ (87GB 225231.0+171747), 0.8\arcdeg\ northeast of \IMP,
to provide an additional check on the stationarity of our other two 
reference sources.  We added this
third reference source after we had gained confidence in the robustness of our
data processing procedures and learned that variations in the observed source
structure of the other reference sources could contribute significantly to our
astrometric uncertainty.
As argued in \pthree\ \citep{GPB-III},
we assume that this third reference source is also extragalactic, even though
its $R=24$ optical counterpart is so faint that no spectrum has yet been obtained for it. 

The cadence of our observation
sessions was determined by a combination of factors.
To ensure that our estimate of proper motion would be only minimally degraded
by the need to also estimate a parallax from the same set of 
astrometric data, we spread the sessions widely across the seasons.
Similarly, to allow us to estimate the orbital contribution to the motion of
the stellar radio emission, we took care that the sessions were also well
distributed in phase with respect to the known binary orbital period of \IMP.
Although the accuracy of the proper-motion estimate might have been improved
by concentrating the sessions at the beginning and the end of the program,
this strategy was effectively precluded by practical considerations regarding
the scheduling and analysis of the observations, not to mention the
indeterminacy in 1997 of the year of launch (2004) of the \GPB\ spacecraft, which was
several years later than the date intended in 1997.  During the last year of the 
program, on the other hand, the end of the
spacecraft operations could be predicted to within an uncertainty of a very few
months.  Our VLBI observations 
that year were scheduled on dates which we calculated
would lead, once all the anticipated VLBI position
measurements were made and analyzed, to relatively high accuracy astrometric 
parameter estimates and low correlations among those estimates.

In addition to the positions from the 35 sessions scheduled in support of
\GPB, we had four reliable positions obtained between 1991 and 1994 
at the same wavelength
by one of us (J.-F. L.) in support of the {\em Hipparcos} mission
\citep{Lestrade+1995}.  As described in \citetalias{GPB-I}, the existence of these VLBI
data also played a role in the selection of \IMP\ as the \GPB\ guide star.
These earlier observations differed from the others in several ways.  In each of
the earlier sessions, only four VLBI stations, all on the United States
mainland, were used, and the observations of \IMP\ were interleaved with 
observations of only \C.  The lower sensitivity of the array used for these sessions
likely explains why four other similar sessions 
between 1992 and 1994 yielded either no 
detection of \IMP, or in one case a relatively weak detection for which no reliable
position could be derived.  Although the four successful sessions greatly
extend our time base, we use the resulting positions with caution, since
they might be affected by different measurement errors and possibly a
different measurement bias than are the later observations. 
These four observations are, however, very valuable in addressing the issue
of a possible third component in the \IMP\ system. 

\section{Position Determinations}
\label{spos}
\subsection{Definition of the Stellar Radio Position}
\label{ssdefpos}

\begin{figure}[tp]
\centering

\includegraphics[height=2.1in,trim=0.0in 0.0in 0.0in 0.0in,clip,angle=00]
{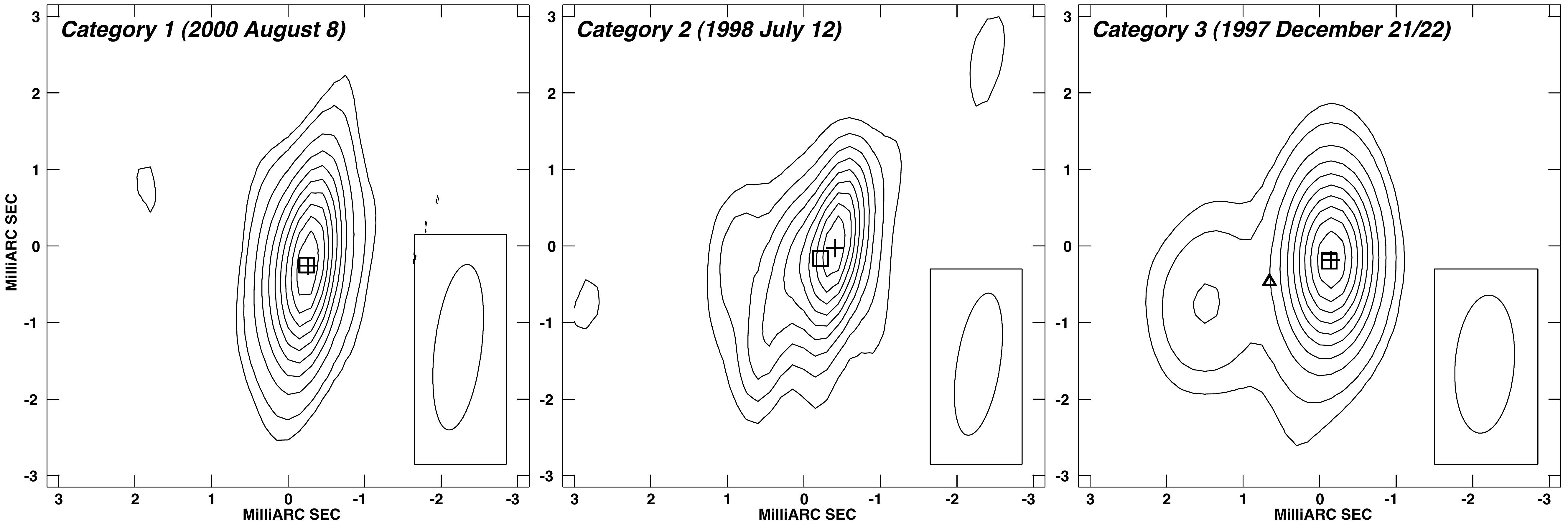}
\caption{
Selection of (Stokes-I, 8.4~GHz) images of \IMP\  
illustrating the three categories of source structures
(see \S~\ref{ssdefpos}).Ê The epoch to 
which each image corresponds is given in parentheses.Ê The full set of 
images of \IMP\  is presented in 
\citetalias{GPB-VII}.  The brightness peak in 
each image is 32.7~mJy/beam (2000 August~8), 0.72~mJy/beam (1998 
July~12), and 31.9~mJy/beam (1997 December~21/22).Ê The contour levels 
displayed in the images for 2000 August~8 and 1997 December~21/22 are 
$-5$ (dotted), 5, 10, 20, 30, 40, 50, 60, 70, 80, and 90\% of the peak 
brightness for that image.Ê The contour levels displayed in the image 
for 1998 July~12 are the same, except that the lowest contour is 
10\%. The restoring beam is shown in the bottom right-hand 
corner of each image.Ê North is up and east is to 
the left.Ê The small cross in each image represents the position of 
the brightness peak.Ê The small open square in each image represents 
the position of the peak of a single elliptical Gaussian fit to a 
region about the source (and down to the noise level of the image). 
The small open triangle in the image for 1997 December~21/22 is the 
position of the midpoint between the eastern and western peaks.
}  
\label{IMimages} \end{figure}

As described in \pfour\ \citep{GPB-IV}, we estimate the position of
the radio emission of \IMP\ at each epoch by a nonstandard, multistep
process that includes both phase-referenced mapping and phase-delay
fitting with a Kalman-filter estimator.  The last major computational
task in this process is to produce a phase-referenced map of \IMP\
based on the final phase calibrations computed with the Kalman-filter
estimator (see \pfour).  
These maps are 256 $\times$ 256 pixels, with pixel size 0.15 mas.
By construction, we know with acceptable
error the coordinate offset of each pixel of this image from the
position of our chosen quasi-inertial point in \C.  Nevertheless, it
is not self-evident which position in each image of \IMP\ should be
used for our subsequent task of estimating \IMP's proper motion from
the full set of positions, because the stellar radio emission is
always detectably extended on the sky.  Our phase-referenced images
reveal this radio emission to fall into three general categories for
the radio source structure of \IMP: (1)~single-peaked with peak
located near the center of a marginally extended source;
(2)~single-peaked with peak located noticeably off center of an
elongated source; and (3)~double-peaked (or in one case apparently
triple-peaked) with maximum separation between peaks of
$\sim$1.5~mas.Ê We show an example from each of these categories in
Figure~\ref{IMimages}.Ê See \pseven\ \citep{GPB-VII} for the complete
set of images and a discussion of them.  Given that the stellar radio
emission was detectably extended, we attempted to find some feature or
pattern in our images with an unambiguous spatial relationship to any
other physically meaningful location in the rotating binary system.
Unfortunately, we were unable to find any such unambiguous
relationship.  We therefore decided to fit an orbit to the ``center"
of the radio emission defined as follows: For each of the sessions for
which the radio image of \IMP\ contained only one, centered, clearly
detected local maximum (category~1, Figure~\ref{IMimages}), we took
the center of the two-dimensional Gaussian component obtained by
fitting such a component (using AIPS task IMFIT or JMFIT) to the 
brightest part of the image.  For
category~2 epochs, we considered two choices. 
The first is to interpolate the brightness-peak position directly from
the image, performing a quadratic interpolation between the pixel
brightnesses (via MAXFIT in AIPS) 
to obtain sub-pixel accuracy on the position of the
brightness peak.  The second is to use the position of the peak of a
Gaussian component fit to the image (see Figure~\ref{IMimages}). 
For category~3 epochs, we considered three choices, namely, the
overall brightness peak, the peak of a single Gaussian component fit
to the entire region of detected emission, and the midpoint between or
among the two or three local brightness maxima (see
Figure~\ref{IMimages}).Ê (For category~1 epochs, the difference
between the position of the interpolated brightness peak and the peak
of the fit Gaussian component is in all cases $<$0.07~mas, and in most
cases $<$0.03~mas, and hence insignificant.)Ê We initially reserved
judgment as to the ``best'' choice, and prepared three sets of
positions for our 35 epochs: (i)~the position of the interpolated
brightness peak for each epoch, (ii)~the position of the center of the
single Gaussian for each epoch, and (iii)~the position of the center
of the single Gaussian for each single-peaked epoch and the positions
of the unweighted midpoint for each multiple-peaked epoch.Ê We then
fit the astrometric model described below in \S\ \ref{ssmodel} to each
of the three sets of positions.Ê We obtained the best fit for choice
(iii): The chi-square per degree of freedom for the resulting fit was
20\% lower than that obtained for choice~(ii) and 30\% lower than that
for choice~(i).Ê We therefore adopted set~(iii) as our standard set of
astrometric positions for \IMP.

For each of our VLBI sessions we also had to specify the
sky coordinates for some reference point in our image of \C.  
For all sessions from 1997 onward, we chose 
this point to be the peak of the C1 component near the core of \C.  
In \citetalias{GPB-III}, we find that C1 is stationary 
in a nearly inertial, extragalactic, reference frame.  
For the four earlier epochs, for which our maps have insufficient resolution 
to adequately resolve this component, 
we used the brightness peak of the core
as the reference point, and assumed that its coordinates are offset from 
those of C1 by the average amount of that offset determined for the later epochs
($-$0.26~mas in \RA\ and $-$0.03~mas in \DEC).

\subsection{Estimated Positions and their Errors}
\label{ssestpos}
The \IMP\ positions resulting from this process are given in Table~\ref{tpos}.
The uncertainties of these \IMP\ positions are dominated by poorly
characterized systematic errors, whose size we can estimate from the scatter
in the differential positions we found for our  pair of
reference sources with the smallest separation on the sky,
\Bb\ and \C.  Since the angular separation of even this
pair is about twice that of \IMP\ from \C, this approach should yield
conservative uncertainties.  For the sessions from 1997 through 2005, we 
estimate our uncertainty based on the weighted root-mean-square
(weighted rms) scatter, 0.045~mas in \RA\ and 0.037~mas in \DEC, of
the postfit residuals that we obtained for the twelve (2002-2005) 
positions of \Bb\ with respect to \C\ \citepalias[see][]{GPB-III}.  However,
in light of the noticeably  larger scatter seen in the \Ba\ positions in the years before
\Bb\ was observed, we multiply the rms scatter in the \Bb\
positions by the ratio of the rms residual scatter 
for each coordinate of \Ba\ (with respect to \C) for the entire period 1997-2005 to 
the rms residual scatter found for only the twelve
sessions during which \Bb\ was observed.  The result is an uncertainty of
$\sim$0.06 mas in each coordinate.  
In addition, there is a nearly constant, common error in all these positions, 
due to the uncertainty of the position of C1.  In \citetalias{GPB-III} we show that, from
1998 through our last VLBI observation in 2005, its mean (J2000) coordinates are
\Ra{22}{53}{57}{7479573} (31) \dec{16}{08}{53}{ 561281} (68),
where the SEs for the last two digits are given in parentheses.  
Over the same span, this component was stationary in our extragalactic reference frame to within
our estimated 1-$\sigma$ bounds of 0.046 and 0.056 \masyr, in \RA\ and \DEC, respectively. 
We cannot make a similar inference of the measurement uncertainties in
the four \IMP\  positions we derived from the 1991-1994 VLBI sessions.  Given the
smaller VLBI array used during these sessions and the resulting poorer
angular resolution and $u$-$v$ coverage, and given that only one reference 
source (\C) was observed, the uncertainty for these four
positions could plausibly be up to about twice as large as that of the other
\IMP\ positions.  In any case, no estimate of the 
measurement uncertainty in our \IMP\ positions was used in
the rest of our astrometric analysis, because, as expected, this uncertainty was 
for all sessions much smaller 
than the rms of the seemingly random scatter of our postfit position residuals, 
which is $\sim$0.4~mas in each coordinate (see Table~\ref{tres} and \S~\ref{sssens}).
This scatter is evidently dominated not by the measurement noise, but 
rather by some other seemingly noise-like contribution.

\begin{deluxetable}{l c c c c c c}
\tablecaption{\protect\IMP\ position estimates\tablenotemark{a} \label{tpos}}
\tabletypesize{\scriptsize}
\tablewidth{0pt}
\tablehead{
\colhead {Observation} & {Epoch}  & \multicolumn{2}{c}{Position} 
 & No.\,of Image \\
\colhead{Date} & \colhead{} 
  & \colhead{$\alpha - \Ra{22}{53}{2}{0}$} 
  & \colhead{$\delta - \dec{16}{50}{28}{0}$} 
  & \colhead{Components} \\
\colhead{} & \colhead{MJD\tablenotemark{b}} & \colhead{(s)}
  & \colhead{(\arcsec)} }
\startdata
1991 12 15 & 48605.06 & 0.276990 & 0.51462 & 1 \\
1993 06 22 & 49160.51 & 0.276091 & 0.48065 & 1 \\
1993 09 13 & 49243.25 & 0.274992 & 0.47385 & 1 \\
1994 07 23 & 49556.45 & 0.274324 & 0.45323 & 1 \\
1997 01 16 & 50464.90 & 0.269798 & 0.37361 & 2 \\
1997 01 18 & 50466.89 & 0.269812 & 0.37288 & 1 \\
1997 11 30 & 50782.03 & 0.268379 & 0.35337 & 1 \\
1997 12 21 & 50803.96 & 0.268350 & 0.34949 & 2 \\
1997 12 27 & 50809.96 & 0.268352 & 0.34835 & 2 \\
1998 03 01 & 50873.78 & 0.268689 & 0.34413 & 1 \\
1998 07 12 & 51006.41 & 0.268662 & 0.34443 & 1 \\
1998 08 08 & 51033.35 & 0.268368 & 0.34217 & 1 \\
1998 09 17 & 51073.24 & 0.267747 & 0.33847 & 1 \\
1999 03 13 & 51250.74 & 0.267332 & 0.31730 & 1 \\
1999 05 15 & 51313.57 & 0.267509 & 0.31785 & 2 \\
1999 09 19 & 51440.23 & 0.266273 & 0.30996 & 2 \\
1999 12 09 & 51521.99 & 0.265512 & 0.29712 & 2 \\
2000 05 15 & 51679.56 & 0.266060 & 0.29084 & 1 \\
2000 08 07 & 51763.34 & 0.265419 & 0.28893 & 1 \\
2000 11 06 & 51854.09 & 0.264135 & 0.27355 & 1 \\
2000 11 07 & 51855.01 & 0.264148 & 0.27339 & 1 \\
2001 03 31 & 51999.73 & 0.264494 & 0.26210 & 1 \\
2001 06 29 & 52089.48 & 0.264483 & 0.26372 & 1 \\
2001 10 20 & 52202.05 & 0.262867 & 0.25081 & 1 \\
2001 12 21 & 52264.99 & 0.262528 & 0.24069 & 1 \\
2002 04 14 & 52378.65 & 0.263190 & 0.23659 & 1 \\
2002 07 14 & 52469.40 & 0.262771 & 0.23533 & 1 \\
2002 11 21 & 52599.06 & 0.261148 & 0.21821 & 1 \\
2003 01 26 & 52665.88 & 0.261121 & 0.20940 &   1 \\
2003 05 18 & 52777.55 & 0.261785 & 0.21022 & 2 \\
2003 09 09 & 52891.24 & 0.260541 & 0.20210 & 1 \\
2003 12 06 & 52979.00 & 0.259686 & 0.18744 & 3 \\
2004 03 06 & 53070.76 & 0.260047 & 0.18194 & 2 \\
2004 05 18 & 53143.58 & 0.260287 & 0.18250 & 1 \\
2004 06 26 & 53182.49 & 0.260099 & 0.18041 & 1 \\ 
2004 12 12 & 53351.00 & 0.258124 & 0.15938 & 1 \\
2005 01 15 & 53385.92 & 0.258220 & 0.15548 & 1 \\
2005 05 28 & 53518.45 & 0.258850 & 0.15447 & 1 \\
2005 07 16 & 53567.41 & 0.258498 & 0.15343 & 1 \\
\enddata
\tablenotetext{a}{For our estimates of the uncertainties of these positions,
see text, \S~\ref{ssestpos}.  }
\tablenotetext{b}{Modified Julian Date = Julian Date $ -$ 2400000.5~d}.  
\end{deluxetable}

The largest source of scatter is likely a highly
variable spatial offset between the stellar radio emission and the
center of the primary component of the \IMP\ binary.  The strongest 
evidence for this assertion is that,
for some of those VLBI sessions marked by emission strong enough to be
detectable in single scans, our VLBI astrometry reveals changes in position
of up to $\sim$1~mas occurring in synchrony with changes in the brightness of
the emission \citep{Lebach+1999}.
In addition, it is plausible that the radio emission is both powered and loosely 
confined by the stellar magnetic field \citep{FranciosiniC1995}.
Moreover, the photospheric spot maps derived
from optical spectroscopy \citep[e.g.,][]{Berdyugina+2000} imply that the
stellar magnetic field is highly variable and asymmetric.  It is therefore not
surprising that the peak of the radio emission is displaced from the center of
the star in a seemingly random manner by an amount comparable to the
$\sim$0.6~mas angular radius of the primary.  For further discussion, see
\S~\ref{sserror}, below, and Papers VI 
\citep{GPB-VI} and VII\nocite{GPB-VII}.

Consequently, we assume that the size and possible anisotropy of the
uncertainty of the VLBI-derived \IMP\ positions are best determined from the 
positions themselves.  Since we find no convincing evidence for any systematic variation
in that uncertainty, we assign identical uncertainty to all of
our \IMP\ positions, including those for the pre-1997 sessions.
 
Smaller position errors are contributed by 
(i)~errors in identifying the reference point in our maps of \C\, and 
(ii)~inaccuracies in our astrometric model (including 
the various inputs to that model).  
Some of these errors vary on timescales of months or years 
and thus cause nonnegligible correlations between the errors
of our estimated \IMP\ positions for sessions that were months or even years
apart.  In fact, the correlations of such errors are evident in the tendency of the
estimated relative positions of our reference sources to vary slowly with
time, rather than exhibit a white-noise behavior (see \citetalias{GPB-III}).  
Moreover, these correlations are not unexpected for two reasons.  
First, it is known that the flux density and also the
structure (i.e., shape) of the emitting regions of \C\  each exhibits
strong autocorrelations over many months (see \citetalias{GPB-II}).
Second, the ionospheric total electron content also correlates over several
years, as it rises and falls with the $\sim$11-year sunspot cycle.  The resulting
position error likely also has a nonzero correlation over several years.  
Moreover, as discussed in \citetalias{GPB-III}, the uncertainty in our position estimates 
caused by inaccuracies in our ionosphere models is one of our major
sources of error.
Conservative estimated standard deviations (see \citetalias{GPB-III} for details) 
on the contributions of errors in our modeling of the propagation 
medium---ionosphere plus troposphere---to the position differences among our 
reference sources range up to $\sim$0.1~mas.  Hence, given the smaller 
angular separation between \IMP\ and \C,  we expect up to $\sim$0.02~mas for the 
corresponding error contribution to our \IMP\ positions.
Any positive correlation among the position measurement errors (or the unmodeled
radio position offsets) for epochs separated by up to several years would
prevent the uncertainty of our proper-motion estimate from falling in
proportion to the square-root of the temporal density of our measured \IMP\
positions, as it would for statistically independent position measurements.
Slowly varying position errors could thus increase the SE of our
proper-motion estimate as much as uncorrelated errors that are severalfold
larger.  However, the variations in our estimates (in \citetalias{GPB-III}) of
the relative positions of our reference sources are severalfold smaller than the 
slowly varying component of the postfit residuals of our \IMP\ solutions.  
(For B2252+172 and 3C 454.3, the closest pair, with separation $\sim$1.4\arcdeg\
on the sky, the weighted rms scatter of the position differences is 0.023~mas
 in \RA\ and 0.051~mas in \DEC.) The small size of these variations
implies that neither the correlated position error due to reference source structure
nor that due to the ionosphere likely accounts for the bulk of the systematic component of
our \IMP\ position residuals.  We postpone a more quantitative treatment of correlated
errors to \S~\ref{sserror}, where we discuss the uncertainty of our results.

\section{Astrometric Solutions}
\label{ssol}

\subsection{The Model}
\label{ssmodel}

We use a
conventional weighted-least-squares (WLS) technique to fit a linearized model 
to the \IMP\ position estimates described above.  The parameters required,
in addition to the proper motion of \IMP, are its position at a reference 
epoch, its parallax, and four scalar parameters to specify the projection on the sky 
of its (assumed) zero-eccentricity orbit of accurately known period
--- nine parameters in all.  As discussed below, we 
considered assigning time-dependent SEs to our VLBI position 
estimates, and we searched for evidence of nonuniform scatter in
our residuals.  Nevertheless, since we found no such evidence and had no a priori
reason to expect any significant nonuniformity, we used uniform weighting to
obtain our final estimates.  On the other hand, we found significantly 
different rms values for the \RhA\ and \DEC\ components of our postfit residuals, 
so we allow for such unequal noise levels in the two coordinates and also for a
nonzero correlation between their errors.  That is, we allow for errors characterized 
by an error ellipse of arbitrary axial ratio and orientation on the sky.  
We estimate the required SEs of the \RhA\ and \DEC\ components of our VLBI
positions from the rms value of the postfit residuals in each coordinate,
and we estimate the correlation between the two coordinates of the error from the 
correlation between the same-epoch postfit \RhA\ and \DEC\ residuals,
by the following procedure:
To obtain for each coordinate an unbiased estimate of the rms measurement
noise value, we increase the rms residual value to account for the number of
degrees of freedom taken out of the residuals in the process of estimating our
free parameters.  Since there is no known bias
between the observed same-epoch correlation of the postfit \RhA\ and \DEC\ 
residuals and the same-epoch correlation in the \RhA\ and \DEC\ 
components of the VLBI position noise, for the latter correlation
we adopt the observed correlation without modification.  We
found that a single iteration of this procedure, starting with the residuals
obtained under the assumption of equal and uncorrelated errors in the \RhA\ and \DEC\ 
components, converged to a limiting value (as estimated from an additional iteration)
for each estimated parameter, to within 1\% of its statistical standard error (SSE).  
Thus for our postfit residuals, $\chi^2$ per degree of freedom is unity and, 
consequently, the SSEs yielded by our WLS fits are unbiased estimates,
at least in the approximation that our errors at different epochs are mutually   
independent and identically distributed.

When we state (above) that our model is a ``linearized" rather than ``linear" one, we
are merely acknowledging that the change in \RA\, due to
a given angular offset on the sky, depends (nonlinearly) on the corresponding \DEC.
However, since \DEC\ is already known to a small fraction of an arcsecond, the partial 
derivatives of the model with respect to its parameters do not, in practice, have to be
recomputed iteratively in our estimation software.  Moreover, as described below, 
we can choose our parameterization of the orbital motion of the radio emission 
to ensure that the orbital part of our model is linear, too.  

We use four choices of reference epoch 
for our positions.  Usually, we
choose the reference epoch to approximate the effective midpoint of the
\GPB\ flight mission, but to facilitate comparisons with other results
(see \S~\ref{scomp}) we 
also computed positions at other epochs, namely the midpoint of our VLBI data, 
the reference epoch of the {\em Hipparcos} Catalogue \citep{PerrymanESAshort1997}, 
and J2000.0.  
Regardless of our choice of reference epoch, all of
our calculations and results are obtained and presented in J2000 coordinates.
 
We compute the parallactic offset numerically at each observing epoch using
a numerical ephemeris (PEP740R, J.\ Chandler 1999, priv.\ comm.)
of the heliocentric orbits of the Earth, planets, 
larger asteroids, and Pluto, and the geocentric lunar orbit, too,
though only Jupiter and Saturn actually affect 
the parallactic offset of \IMP\ to a nonnegligible degree.  
The aberration effect of Earth's motion is generally far larger; the largest 
aberration term, the annual aberration, can be as much as $\sim$20\arcsec.  
However, because the models used in VLBI data
processing are always computed in a solar-system barycentric reference
frame, all the known aberration effects are removed from the VLBI data at
an early stage of processing.  At no stage of our analysis do the
ephemerides contribute more than a (negligible) few microarcseconds of 
uncertainty to our position results.

We restrict our model to include only a zero-eccentricity orbit, because our
data set is far less sensitive to eccentricity than is optical spectroscopy,
which bounds any true eccentricity of the \IMP\ binary orbit below $\sim$0.01
\citep{BerdyuginaIT1999, Marsden+2005}. 
Since our 35 VLBI position estimates are characterized by $\sim$0.4~mas rms
noise-like scatter in each coordinate and our WLS orbit of the radio emission
on the sky has a semimajor axis of only $\sim$0.9~mas, the effect of such
small eccentricities is not detectable with our data set.  Moreover, using
software that allows eccentric orbits, developed as part of the {\em Hipparcos}
frame-tie program of \citet{Lestrade+1995}, we confirmed that the improvement
in the fit to our VLBI-determined positions achievable with a grid search over all
possible values of eccentricity and time of periastron passage is not
statistically significant.  In any case, the resulting change in our estimate 
of proper motion would be no more than a negligible 0.01~\masyr\ in either 
coordinate.

Our VLBI position estimates also lack the precision necessary to
detect plausible departures of the orbital period of the radio emission from 
the value derived from optical data.  Even if, to increase the time span of
our data set,  we include (with equal weight) 
the four available VLBI positions from 1991-1994, 
the SSE of our WLS estimate of orbital period 
is $\sim$0.01~d, which is more than a factor of 300 larger than that of
the spectroscopic result of \citet{Marsden+2005}.  Moreover, our WLS 
adjustment to that result is
not statistically significant.  Given the lack of any convincing evidence
of period variation and the lack of any strong reason to believe that the 
source of the radio emission drifts systematically with respect to the line
connecting the two stellar components of the \IMP\ binary, we adopt the 
optical value for the orbital period \citep[24.64877~d;][]{Marsden+2005}.  
We considered the advisability of also adopting optically-derived values for the 
inclination and the position along the orbit.  However, the optically-derived 
inclination has only similar accuracy to ours
(see \S\ \ref{scomp}) and is obtained only indirectly
(by combining radial velocities from spectroscopy with mass estimates based on
results from spectroscopy and photometry).  Furthermore, we cannot rule out the
possibility that some type of interaction between the two components of the
binary causes the peak of the radio emission, on average, to either lag or
precede the position of the primary in its orbit.  We therefore estimate these
two orbital parameters from our VLBI data, without any a priori constraint on their  
values.  

We usually represent the orbital motion in each coordinate by the sum 
of one term proportional to the sine of orbital phase and one proportional to the 
cosine.  The resulting model is then strictly linear in the unknown amplitudes.  
(For the mathematical formulation of these terms, see note {\em e} to Table~\ref{tfinal}.)
The choice of the zero point of orbital phase has no 
effect on the final fit to our VLBI positions or on the values of the 
non-orbital parameters.  To facilitate comparison with the optical results
of Marsden et al., we take as our zero point the epoch that they determined 
to be a (heliocentric) time of conjunction with the primary farther from us
than the secondary: Julian day 2,450,342.905.   
To further aid such comparisons, we can alternatively
parameterize the orbital motion by its semimajor axis, axial ratio
(as projected on the sky), nodal position angle, and time of conjunction.  
Since the orbit model is not a linear function of these four alternate parameters, 
we compute them by iterating their linearized WLS
estimates to convergence.  As in the previously mentioned case of our iterating
to determine the SEs of the \RhA\ and \DEC\ components of 
our VLBI positions, convergence of the orbit parameters is reached after just a 
few iterations.
The orbit on the sky specified by our 
converged WLS estimates of the alternative parameters 
is identical (to within our fully adequate computational precision) to that obtained 
with our model parameterized with separate sinusoidal terms in \RA\ and \DEC.  
Consequently, neither the postfit residuals nor the estimate of any non-orbital 
parameter differs between the two model parameterizations.  Numerical 
confirmation of this lack of change provided us with a useful check on our
WLS fitting software.

Should our model also allow for the possibility that the close binary system is 
orbited by a third, more distant, companion?  One cannot rule out such 
a companion on astrophysical grounds; to the contrary, about half of known
close binaries have more distant stellar companions
\citep[e.g.,][]{MayorM1987,Tokovinin+2006}.  Moreover, the potential impact of
such a companion on the accuracy of our VLBI proper-motion estimate is too
large to ignore.  For example, a star of 1 solar mass separated from the \IMP\
binary by $1''$ on the sky could cause an angular acceleration of the binary
at a rate as high as 0.05~\masyryr, with the maximum acceleration occurring 
when the distance of the star from us is identical to that of the binary.  If 
our estimate of proper motion were used in the mission data analysis without
taking account of such 
a possible acceleration, an unacceptably large error in the \GPB\
relativity tests might be introduced, because the mean epoch of our VLBI data
is $\sim$4 yr earlier than that of the spacecraft gyro measurements.
We addressed this potential problem in several ways.  First, we 
initiated a campaign of optical observations using HST and ground-based
telescopes, to place limits on the maximum brightness of any third companion
to the binary (see \S~\ref{sssens}).  Second, we designed our
program of VLBI observations to meet the \GPB\ accuracy requirement for proper
motion, even were we to solve for a constant acceleration on the sky---what we
call ``proper acceleration"--- along with the other required parameters.  This 
acceleration term would allow us to remove most of the effect on the \GPB\ 
experiment of a companion in an orbit with a period that is too long to be 
clearly identifiable in our postfit residuals.  
Later, during our data analysis, we compared the quality of our
fits to our VLBI-determined positions with and without adding
proper acceleration to our model.  As described below in \S~\ref{sserror}, we
concluded that retention of this term is not justified.  (Admittedly, an extremely
eccentric orbit whose periastron passage occurred during the span of our VLBI
data would be modeled poorly in either case, but such orbits and timing are 
a priori unlikely.)  Finally, we visually inspected our postfit residuals and also
performed a periodogram analysis of them.  We found no clear indication of a
periodic component, and hence no need to incorporate a second orbit in our
model.

\subsection{Sensitivity of the Results to Various Analysis Options}
\label{sssens}

\begin{figure}[tp]
\centering
\includegraphics[height=6.5in,trim=0.7in 1.0in 1in 0.0in,clip]
{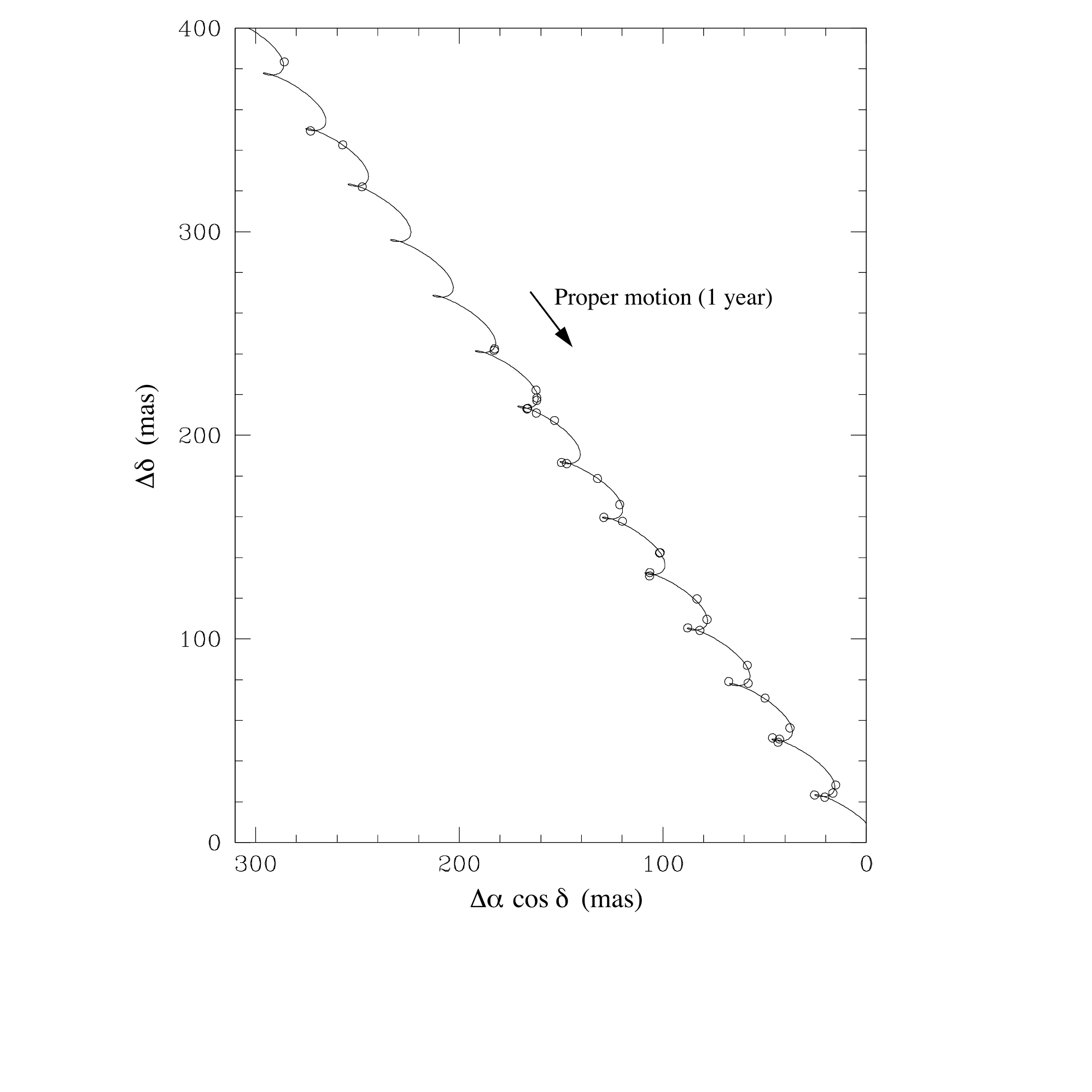}
\caption{
The 39 VLBI-derived positions of \IMP\ in Table~\ref{tpos}
plotted to scale on the sky
(with north at the top and east to the left).  In this plot only,
the coordinate origin is arbitrary.  At this scale, our $\sim$$0.1$~mas 
estimated measurement errors (see \S~\ref{ssestpos}) are far smaller 
than the diameter of the position symbols.
The plotted curve shows the astrometric fit of our chosen Solution~1 of 
Table~\ref{tres}, except that the orbital terms, only marginally visible at 
the scale of the figure, are excluded for clarity.  At this scale, all
data appear to be well fit, even though the first four data 
points (at upper left), spanning the years 1991 to 1994, 
were given no weight in this particular solution.
}  
\label{fimpsky} \end{figure}

\begin{figure}[tp]
\centering
\includegraphics[height=5.9in,trim=0 0.0in 0 1.1in,clip]
{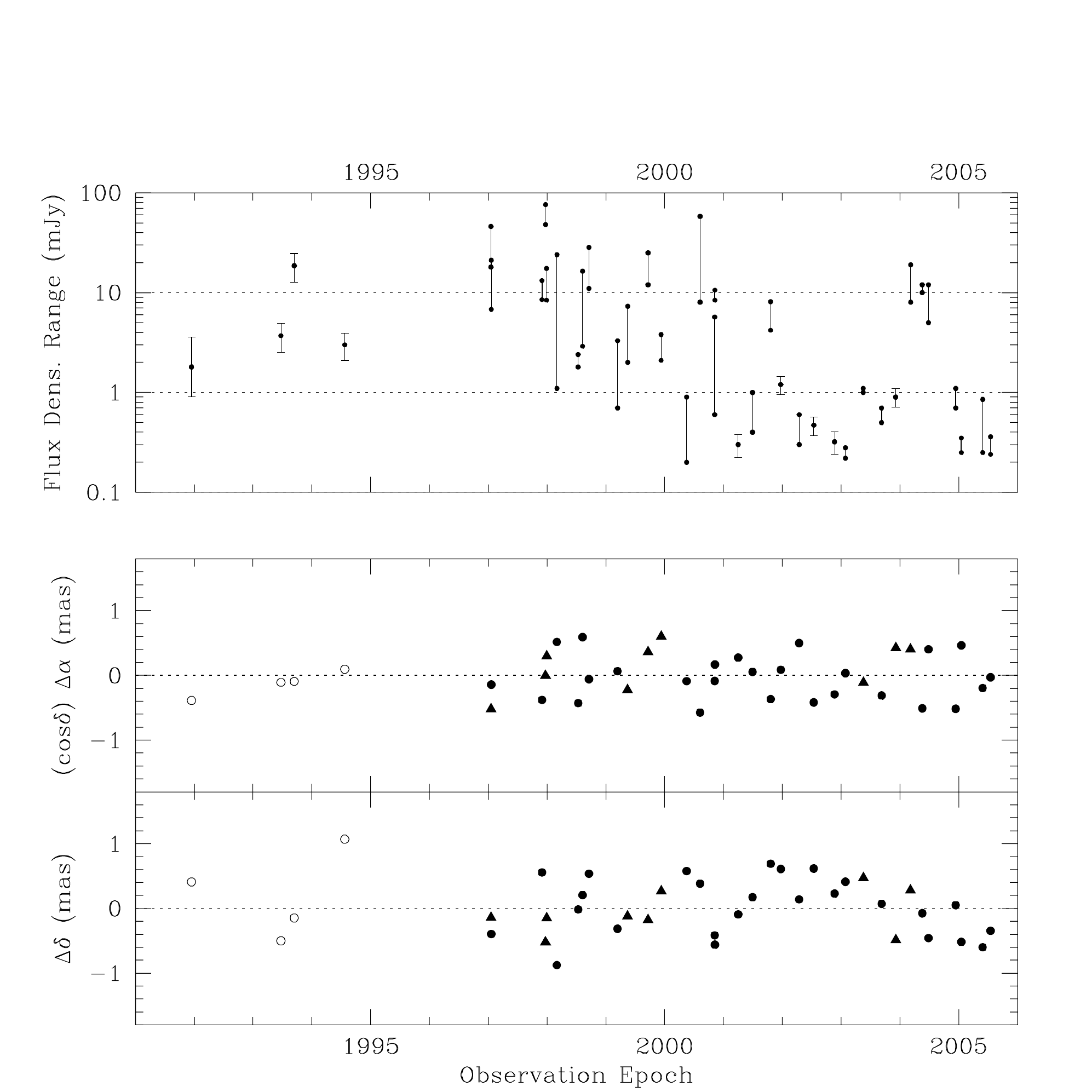}
\caption{
The upper panel shows flux densities for our 39 VLBI sessions.  For 30 of 
these sessions, we show the range between the highest and
lowest values of the flux density observed at the VLA, after smoothing with a
boxcar window about 20 minutes wide.  For the 2001 March session, there was
insufficient signal at the VLA to clearly detect any change in the flux
density, and so we plot just a single value.  For the eight other sessions, 
the VLA was
not used; for these sessions we also plot only a single flux density, the
total flux density contained in our VLBI image of the emission.  For each
of these nine sessions, error bars indicate our estimated 70\%-confidence
interval for the total flux density, with allowance for amplitude calibration
errors and the noise levels in our VLBI images.  The two lower panels show 
the position residuals for all sessions for Solution~1 in Table~\ref{tres} 
(see also Table~\ref{tfinal}).  We plot
unweighted points as open circles and weighted points as either closed circles
or triangles, with each triangle indicating a position computed as the mean
position of the two or three resolved peaks in the stellar radio image for
that session (see text, \S~\ref{ssdefpos}).
We  plot no error bars since we have no valid basis for assigning 
effective SEs to our VLBI position estimates for \IMP, other than
the scatter shown in this figure.
}  
\label{fresvt} \end{figure}

In Table~\ref{tres}, we present the proper-motion and parallax estimates from
seven different WLS astrometric solutions,
which taken together justify our reliance upon the first one for our final results.  
We obtained this Solution~1 by fitting our nine-parameter model
(without proper acceleration) to the \IMP\ positions we determined for all
35 VLBI sessions scheduled in support of \GPB.  A plot of the fit of this
solution to all the position data is shown in Figure~\ref{fimpsky}. 
The corresponding postfit residuals are plotted in Figure~\ref{fresvt}.  
Solution~2 differs from Solution~1 in that we also estimated the \RhA\
and \DEC\ components of a constant proper acceleration.  Solutions 3 and 4
are the two corresponding sets of results obtained by adding to our
primary data set the position estimates we derived from our own reduction of
the four sessions of \IMP\ data obtained by \citet{Lestrade+1995}.  
Solution~5 differs from Solution~1 in that we excluded the nine 
positions derived from those VLBI images in which the radio
emission from \IMP\ was clearly resolved into more than one component.  
For Solution~6 we fit our nine-parameter model to the set of 
positions estimated using only AIPS processing (as opposed to 
``MA" processing, as described in \citetalias{GPB-IV} and discussed below) 
in the phase-referenced mapping steps of our data analysis.  We now explain
why these results lead us to conclude that the first set, Solution~1, provides
reasonable estimates of \IMP's parameters.  The motivation for Solution~7 we
defer until \S~\ref{ssseeec}.

\begin{deluxetable}{l@{ } c@{ } c@{\hspace{0.06in} }
        c@{ } c@{ } c@{ } c@{ } c@{ } c@{ } 
        c@{ } c@{ } c@{ } c@{ } c@{ } }
\tablecaption{Comparison of 
astrometric solutions\tablenotemark{a}\label{tres}}
\tabletypesize{\scriptsize}
\tablewidth{0pt}
\tablehead{
Solution 
& \multicolumn{2}{c}{Proper Motion} & \colhead{Parallax}
& \multicolumn{2}{c}{Proper Acceleration} 
& \multicolumn{2}{c}{RMS Residuals}    \\
\colhead{ }  
  & \colhead{$\mua + 20.83$}  & \colhead{$\mud + 27.27$}  
  & \colhead{$\pi - 10.37$} & \colhead{\dotmua}  & \colhead{\dotmud}
  & \colhead{$\alpha$}  & \colhead{$\delta$}     \\
  & \colhead{(\masyr)}  & \colhead{(\masyr)}  & \colhead{(mas)}
  & \colhead{(\masyryr)}  & \colhead{(\masyryr)}
  & \colhead{(mas)} & \colhead{(mas)} }
\startdata
1. Chosen\tablenotemark{b}              
 & $\phantom{-}$0.00 $\pm$ 0.03  & 
     $\phantom{-}$0.00 $\pm$ 0.03  & $\phantom{-}$0.00 $\pm$ 0.07 & \nodata & \nodata
 & 0.354 & 0.416  \\
2. With proper acceleration\tablenotemark{b}            
 & $-$0.07 $\pm$ 0.10  & $-$0.31 $\pm$ 0.10  & $-$0.03 $\pm$ 0.07 
 & $-$0.018 $\pm$ 0.025  &  $-$0.085 $\pm$ 0.025
 & 0.349 & 0.357  \\
3. With the 4 early epochs\tablenotemark{b}       
 & +0.01 $\pm$ 0.02  & $-$0.01 $\pm$ 0.02  & +0.01 $\pm$ 0.07 & \nodata & \nodata
 & 0.340 & 0.440  \\
4. With 4 early epochs \& accel.\tablenotemark{b} 
 & $-$0.03 $\pm$ 0.05  & $-$0.04 $\pm$ 0.06  & +0.01 $\pm$ 0.07
 & $-$0.008 $\pm$ 0.008 & $-$0.006 $\pm$ 0.011 
 & 0.336 & 0.437  \\
5. Without multi-comp.\ epochs\tablenotemark{b}\tablenotemark{c} 
 &  $-$0.02 $\pm$ 0.03  & $-$0.02 $\pm$ 0.04  & +0.06 $\pm$ 0.08 & \nodata & \nodata
 & 0.333 & 0.434  \\
6. Using ``AIPS-only" positions\tablenotemark{b} 
 &  $-$0.05 $\pm$ 0.03  & +0.05 $\pm$ 0.03  & +0.07 $\pm$ 0.08 & \nodata & \nodata
 & 0.394 & 0.446  \\
7. Without last 5 epochs\tablenotemark{d} 
 &  $\phantom{-}$0.00 $\pm$ 0.03 
 & +0.07 $\pm$ 0.04  & +0.02 $\pm$ 0.08 & \nodata & \nodata
 & 0.349 & 0.383  \\
\enddata
\tablenotetext{a}{Here and elsewhere, the \RhA\  component  of proper motion, \mua,
and its time derivative, \dotmua,
have been multiplied by the factor $\cos\delta$, so that they are, respectively, 
the rates of motion and acceleration on the sky,  i.e., 
$ \mu_{\alpha*} = \mu_{\alpha}\cos\delta$,
where $\mu_{\alpha}$ is the time derivative of  right ascension, $\alpha$.
The errors shown are the SSEs yielded for each 
parameter by the WLS fits.
Throughout this paper, we employ J2000 coordinates.  For Solutions 2 and 4,
the tabulated proper motion is for epoch JD 2453403.0 (2005 Feb 1, $\sim$2005.08), 
the approximate midpoint of the \GPB\ science data.}
\tablenotetext{b}{See text, \S~\ref{sssens}.}
\tablenotetext{c}{This solution was obtained after excluding the nine epochs 
for which the stellar radio source exhibited more than one brightness peak
(see Table \ref{tpos}).}
\tablenotetext{d}{See text, \S~\ref{ssseeec}.}
\end{deluxetable}

Comparison of the first two solutions in Table~\ref{tres} reveals that
inclusion of the acceleration term has at least two important effects.  First,
it increases the SSE of each of the two components of proper motion, \mua\ and
\mud, between three- and fourfold.  The main cause of this increase is readily
understood:  Because the 
2005.08 reference epoch of the (time-varying) proper motion
is less than half a year before the end of our $\sim$8.5~yr data span, 
the estimates of \mua\ and \mud\ are, respectively, highly correlated
(96\%) with those of \dotmua\ and \dotmud.  The increased SSEs
follow from these high correlations.  
Second, inclusion of the acceleration causes changes in the proper-motion estimate.
Indeed, the $\sim$0.3~\masyr\ ($\sim$3-$\sigma$) 
change in \mud\ exceeds the nominal accuracy requirement of
our VLBI program.  Therefore, we next justify why the proper-motion estimate 
from Solution~1 is clearly more appropriate for \GPB\ than is the one
from Solution~2.

Key support for this judgment is provided by Solutions 3 and 4, which differ
from Solutions 1 and 2 in that we added to our set of \IMP\ positions the four we 
determined from the VLBI observations of \citet{Lestrade+1995}.
Comparison of Solution~3 with Solution~1 reveals that, while the
SSEs of \mua\ and \mud\ decrease by roughly one-third (as might be
expected as a result of the 50\% longer time span of the larger data set), the
changes in the WLS estimates of \mua\ and \mud\ are each smaller than their
respective SSEs in either of these solutions.  In this sense the estimates 
from Solutions~1 and 3 are consistent with each other.  On the other hand,
comparison of Solution~4 with Solution~2 
reveals that not only do the added VLBI epochs reduce the
SSEs of the proper-motion estimate, but they also significantly alter that
estimate, pushing it back toward that of Solution~1.  The 
(previously) worrisome change in 
the estimate of \mud\ introduced by solving for proper
acceleration is reduced by a factor of ten to a value much smaller than our
nominal accuracy requirement.  The high correlation between the
proper-motion and proper-acceleration estimates (94\% in Solution~4)
assures that there is
also a corresponding effect on the proper-acceleration estimates.
In fact, in Solution~4 neither \dotmua\ nor \dotmud\ is significantly
different from zero.  In addition, the model from Solution~2 
fails to fit the (unweighted) 1991-1994 data, with its postfit \DEC\ residuals 
being up to 4~mas, which is about ten times the rms postfit scatter in 
declination of the weighted points.  In contrast, the model from Solution~1
fits the unweighted early data 
to within $\sim$1~mas, as shown in the residual plot (Figure~\ref{fresvt}).

We draw three conclusions from our comparisons of these solutions:
(i) Solution~2 gives a poor representation of the motion of \IMP; 
(ii) adoption of Solution~3
rather than Solution~1 as our nominal solution would have little effect on our
proper-motion estimate, while decreasing its SSE by roughly 30\%; and   
(iii) Solution~4 is less credible than Solution~1, since there is
no independent evidence for nonzero \dotmua\ and \dotmud.  Furthermore, with
other colleagues, we carried out an extensive, multifaceted observational
search for a third stellar component.  We found no credible evidence of any
such,\footnote{The  observational bound on the maximum optical 
brightness of undetected companions is, of course, wavelength dependent
and also strongly dependent on angular separation.  For example, based on
our HST observations obtained with the 
1042~nm WFPC2 filter, the minimum magnitude differences between
the unresolved IM Peg binary and any third companion at 
angular separations on the sky of 0\farcs{1}, 0\farcs{5}, 1\arcsec, and 5\arcsec\
are, respectively, about 5, 9, 11, and 16 magnitudes.
Observations through the 334~nm WFPC2 filter, as well as a 
wide variety of ground-based observations, listed in \citetalias{GPB-I},
yielded other useful magnitude limits, applicable to both larger and smaller 
angular separations.}
allowing us to infer,  under reasonable assumptions, that the 
probability of a detectable nonzero acceleration due to a third stellar 
component is less than $\sim$5\%.  While the observational
bounds on companions are difficult to quantify and summarize, this low
probability is one of the computational results of a Bayesian statistical study
(J.~Chandler 2007, priv. comm.) that combined these bounds with a range of
plausible prior distributions of hypothetical third components with respect to mass,
orbit parameters, luminosity, etc.  Specifically, in light of the observational 
bounds on any companion of \IMP, unless more than 80\% of RS~CVn
binaries are assumed to have companions, there is a probability of only
$\sim$5\% that \IMP\ in particular has one that causes a proper acceleration
of the binary detectable at even the 1-$\sigma$ level in Solution~4.  (The
separate principal result of this study is a $\sim$95\%-confidence statistical 
inference that the error in the guiding behavior of the \GPB\ spacecraft 
caused by light from any third component is a negligible source of error,
$<$0.006~\masyr, for the mission.)  

Table \ref{tres} also contains the results of two additional solutions made
to test the sensitivity of our results to each of two other choices 
we made.  As noted in \S~\ref{ssdefpos} and Table \ref{tres}, 
among the  35 VLBI position estimates that we used to obtain 
Solution~1 are nine from epochs at which our
images of \IMP\ resolve its radio emission into two or, in one instance, three
components, separated by 1 to 2~mas (see Table~\ref{tpos}; see 
also Figure~2 in \citetalias{GPB-VII} for images).
Although the postfit residuals are 
neither noticeably larger or smaller at these nine than at the other epochs, we
naturally were concerned that the VLBI positions at these epochs might
be subject to systematic differences from those at the other epochs, in such
a way as to significantly affect our results.  We thus made Solution~5, with 
these nine epochs of data removed.  Comparison of the 
estimates from this solution with those from Solution~1 is reassuring:  
Our proper-motion estimate changes by no more than 0.02~\masyr\
in either coordinate.  
Moreover, there is no consistent change in the scatter of our postfit
residuals:  The rms of the \RhA\ 
residuals decreases 6\%, but the rms in \DEC\ increases 4\%.  
Given these results, we conclude that Solution~1 need not be modified to
account for the stellar radio emission's being at times resolved into 
spatially separated components.   Solution~1 is also the most consistent
with our general preference for the inclusion
of all available high quality data, with uniform weighting of all sessions, 
since no other weighting is clearly justified.
Of course, we could have exercised this preference still further by adopting
Solution~3, obtained by fully weighting the four VLBI-determined positions 
available from 1991 to 1994.  We did not do so for fear of underestimating 
the uncertainties associated with these positions.  The smaller, more compact
VLBI array used at these epochs makes the uncertainty of the corresponding
position estimates 
larger and also more difficult to estimate.  A conservative doubling of
the position SEs, together with allowance for a plausible unknown common 
bias to these four positions relative to the others, 
would yield parameter estimates and SSEs closer
to those of Solution~1 than to those of Solution~3.  For simplicity, we choose
to rely on Solution~1.

Solution~6 in Table \ref{tres} was made to explore a 
different aspect of our analysis, namely our use, as mentioned  in 
\S~\ref{ssdefpos}, of VLBI phase calibrations
derived from a Kalman-filter analysis of our VLBI data, rather than those 
computed within the basic AIPS package.
To improve the accuracy of the phase models we produced with our
Kalman-filter analysis, we normally included an ionospheric-delay model and 
employed updated, more accurate, Earth-orientation parameters 
and antenna positions than those originally contained in the AIPS data files.
For Solution 6, we instead relied on AIPS calibrations alone.
Consequently, for each parameter
the size of the difference between the results from Solutions 1 and 6
provides a conservative, if rough,
indication of the former result's uncertainty due to inaccuracies in
our data-reduction models, or at least due to  those inaccuracies that are not
common to both of our phase-calibration procedures.
Once again, the comparison between Solutions 1 and 6 is 
somewhat reassuring:  Our proper-motion
estimate changes by no more than 0.05~\masyr\ in either coordinate.

\subsection{Error Analysis}
\label{sserror}

\subsubsection{Postfit residuals}
\label{ssspfres}

For each solution in Table \ref{tres}, the tabulated SSEs are those
obtained directly from our WLS analysis. 
As discussed in \S~\ref{ssmodel}, these SSEs would be unbiased estimates of 
the true SEs if all the errors in our measured positions were  
accounted for by our weighting scheme, which is based on the 
approximation that the position errors at all measurement epochs 
have identical and independent distributions.  In 
this context, we regard any noise-like contribution to the offset on the 
sky between the position of the radio emission and that of the center
of the primary as being a contribution to the measurement error.
The causes of these (varying) offsets are not known, and so the best
checks on their statistical properties are provided by 
examining our postfit residuals.

At first glance, the residuals in Figure~\ref{fresvt}
look like white noise, but there is a strong
suggestion that the \DEC\ residuals, at least for the later years, are not
statistically independent between epochs separated by less than two years.
Since such autocorrelation would effectively reduce the number of independent
measurements in our data set, it would tend to increase the true SSEs of our
parameter estimates, and hence it is potentially important.  We return to this
issue below, after we complete our overview of the plots of our residuals.

\begin{figure}[tp]

\centering
\includegraphics[height=3.8in,trim=0 0.0in 0 3.5in,clip]
{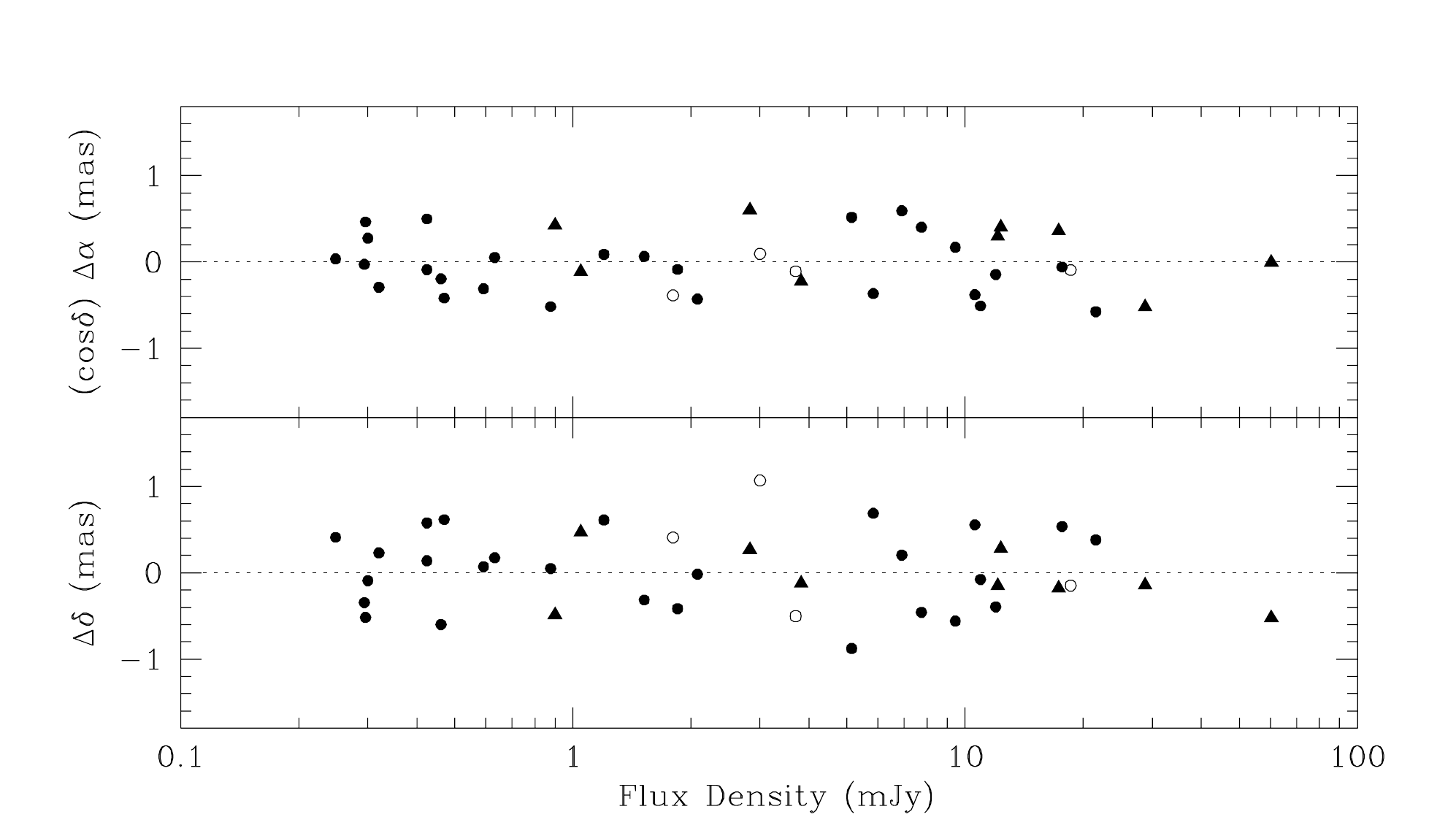}
\caption{
The same residuals and symbols as in Figure~\ref{fresvt}, but plotted against 
\IMP's flux density, with a logarithmic scale.  For each of those sessions 
for which a highest and a lowest value of flux density
are displayed in the top panel of Figure~\ref{fresvt}, the geometric mean
of those two values is used here.  The residuals exhibit no 
systematic dependence on flux density.
}  
\label{fresvfd} \end{figure}

The top panel of Figure~\ref{fresvt} indicates the range of stellar flux
density detected at the VLA during each VLBI session when that instrument
observed \IMP.  Although the stellar flux density varied by a factor of ten or
more during most years of our observations, it also exhibited an unmistakable
downward trend over the 8.5 years of our observations.
Consequently, if the variance of our position measurement errors (or any
systematic bias in those measurements) were strongly correlated with flux
density, our WLS proper-motion estimate would be subject to additional
uncertainty that is not accounted for in our tabulated SSEs.
To explore this possibility, we plot in Figure~\ref{fresvfd} our
position residuals against flux density.
No noticeable trend is seen, which suggests
that any correlation between position residual and flux density is small and
likely merely the result of random fluctuations within our modest-sized sample
of measurements.  There is also no evident relation between the
amplitude of the scatter of the residuals and the measured flux density.
Specifically, there is neither a significant linear trend nor any indication
of larger than usual residuals being associated with either the highest or
lowest flux densities observed.  These observations bolster our decision to
weight all our measured positions equally, regardless of the stellar flux
density during the various sessions.

\begin{figure}[tp]
\centering
\includegraphics[height=3.8in,trim=0 0.0in 0 3.5in,clip]
{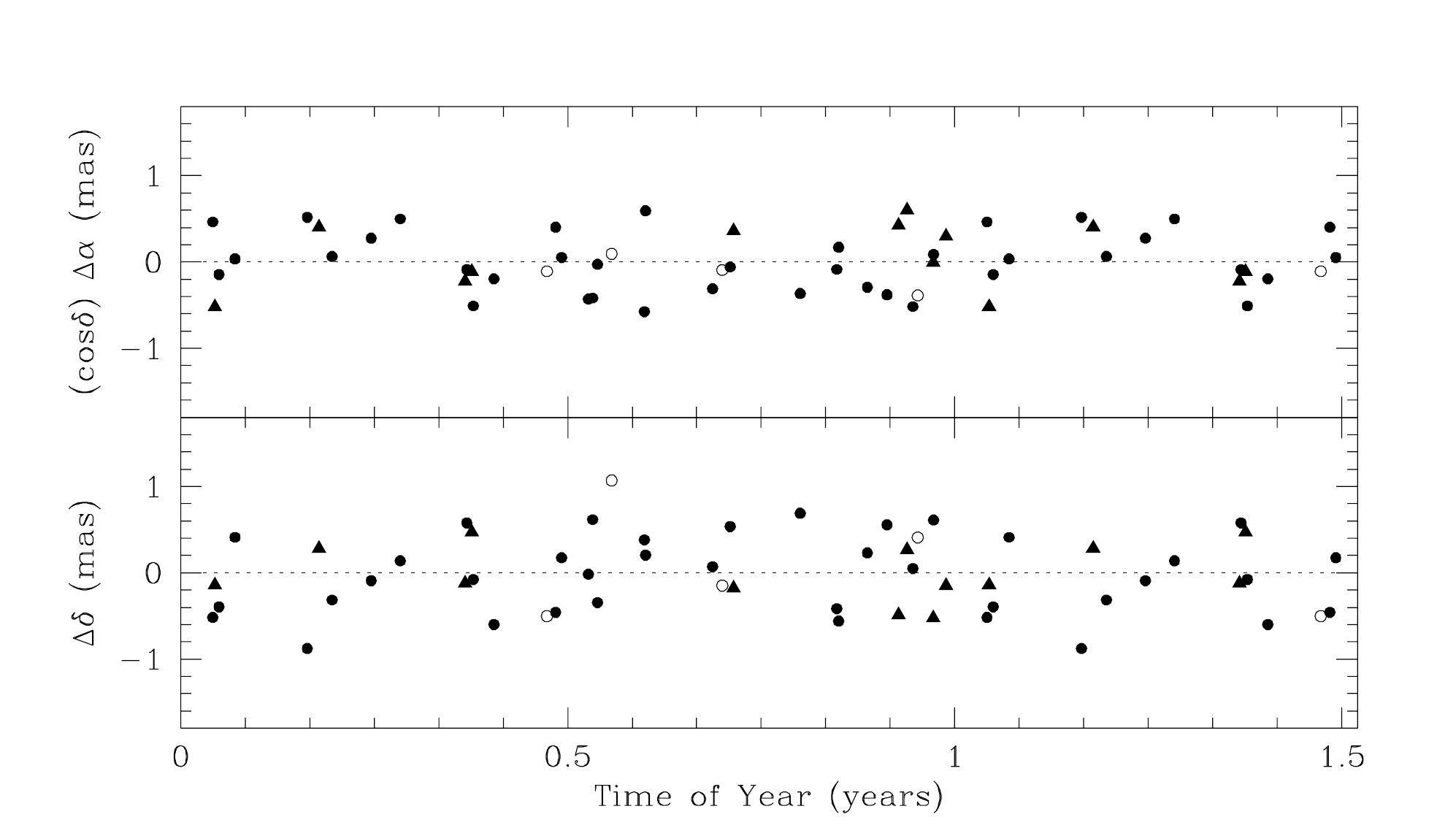}
\caption{
The same residuals and symbols as in Figure~\ref{fresvt}, but plotted against 
time of year in fractional years from J2000.0.  To better demonstrate the 
lack of any trend from fall through winter, all residuals from the first half
of any year are also plotted a second time shifted to the right by one year.
}  
\label{fresvyr} \end{figure}

Similarly, our parallax estimate and its true SE could be
adversely affected were the errors in the VLBI position measurements 
seasonally dependent.  We therefore plot our postfit residuals against time 
of year in Figure~\ref{fresvyr}.  Here, too, the residuals
appear to be merely noise with constant variance, with the possible exception
of two one- or two-month-long parts of the year 
with several neighboring  \RhA\ residuals of the same sign.
We conclude that the residuals do not justify increasing 
the SSE of our WLS estimate of \IMP's parallax.

\begin{figure}[tp]
\centering
\includegraphics[height=3.8in,trim=0 0.0in 0 3.5in,clip]
{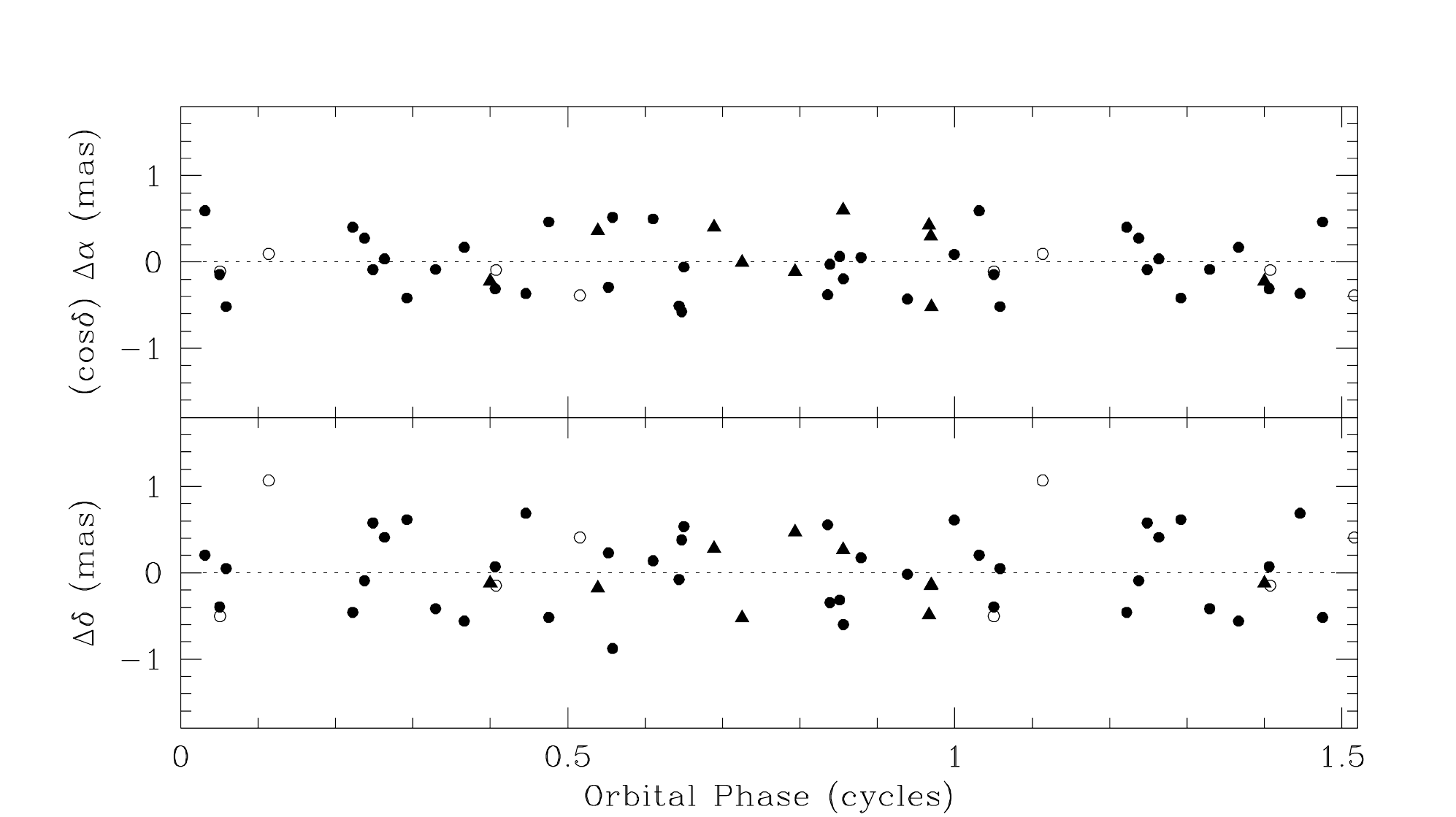}
\caption{
The same residuals and symbols as in Figure~\ref{fresvt}, but plotted against 
binary orbit phase in cycles.  To better demonstrate the 
lack of any trend near the cycle boundary, all residuals from the first half
of any cycle are also plotted a second time shifted to the right by one 
cycle.
}  
\label{fresvp} \end{figure}

Just as the accuracy of our parallax estimate could be affected by a 
seasonal dependence in our VLBI position errors, our estimates of the orbit 
terms and their true SEs
could be affected were our position errors dependent on orbit phase.  
Figure~\ref{fresvp} shows our postfit position residuals
plotted against orbit phase.  Yet again, the
residuals look like noise, with no systematic dependence on phase.  There
appears, therefore,
to be no need to add any additional terms to our orbit model, or to
allow for phase dependence of our measurement uncertainty.  
(Similarly, the plot of flux density against orbital phase, which is
presented in \citetalias{GPB-VII}, shows no credible systematic
relation between flux density and orbital phase.)
We also computed Lomb-Scargle periodograms \citep{Press+1992}
of our residuals.  (For unevenly sampled time series like ours, such 
periodograms are more useful than Fourier transforms.)
No peaks in the periodogram stood out as obviously significant.  However, 
of the three (roughly comparable) highest peaks in the \DEC\ periodogram,
the highest one (with a semi-amplitude corresponding to $\sim$0.36 mas) 
occurred at period 8.16~d, which is within 0.06~d of exactly one-third of 
\IMP's orbital period.  It is unclear if this peak is 
significant, especially since there is no 
corresponding power excess in the \RhA\ residuals.  On the other hand, were 
the residuals white noise, the probability that the 
highest peak in the periodogram
of either the \RhA\ or the \DEC\ residuals would fall so close in frequency
to either exactly two times or three times the reciprocal of the $\sim$24.65~d 
orbit period is only $\sim$0.007.  Thus the existence of the peak in the 
periodogram at 8.16 d suggests that the residuals 
contain at least some quasiperiodic signal.  This suggestion 
is consistent with our understanding (see \S~\ref{ssestpos}) that the 
residuals are largely due to
variable offsets of the peak of the radio emission from the center of 
\IMP's chromospherically active primary.  Specifically, such offsets are 
likely related to the structure of the stellar magnetic field, which 
is, in turn, likely related to the photospheric spot distribution.  Since we know 
from optical spectroscopy that the spot distribution varies only 
slowly, i.e., on timescales of months and years in a coordinate system
that rotates with the same period as the near-circular orbit of the IM Peg
binary \citep{Berdyugina+2000}, 
presumably as a result of tidal locking,
the offsets might well have a quasiperiodic component.  Nevertheless, 
given the weakness of the statistical evidence for any periodic component
in our residuals, we see no need to adjust 
the uncertainties of either our measured positions or the resulting 
parameter estimates to account for a possible dependence of those 
measurements on orbit phase.

In any case, the ``orbit" we determine for the radio
emission does not necessarily provide unbiased estimates of the size, shape, 
and phase of the orbit of the \IMP\ primary.  
We discuss the strength of the correspondence in \S~\ref{scomp} and 
\citetalias{GPB-VI}, on account of its astrophysical  significance.  
Such a correspondence, however, is not needed to meet the needs 
of the \GPB\ mission.  Solving for the orbital motion of the radio 
emission is a reasonable and demonstrably effective way of improving our fit 
to our measured positions and arguably, therefore, also the accuracy of our
proper-motion and parallax estimates.  

Could a non-Gaussian distribution of our position noise significantly
affect the accuracy of our parameter estimates? 
Since this noise is in large part due to offsets of the 
peak of the stellar radio emission from the center of the 
\IMP\ primary, or from
some other nonstochastic mean position, there is no good reason to believe
that this distribution is Gaussian.  Nevertheless, because each of our
WLS parameter estimates is calculated as a linear
combination of the (2-dimensional) prefit residuals of our 
35 measured positions with respect to our ``initial-guess" 
model, the statistical distribution of each estimate will 
be approximately Gaussian, since each 70-term summation comes reasonably
close to satisfying the conditions of the central limit theorems.  
Moreover, the SSEs of the parameter estimates yielded by the WLS 
approach do not depend on the shape of the measurement error distribution,
but only on the measurement SEs.

\subsubsection{Epoch-to-epoch error correlation}
\label{ssseeec}

What about the apparently nonzero autocorrelation in the \DEC\ residuals, 
over time lags up to one or two years?  The \DEC\ residuals in 
Figure~\ref{fresvt} do not look like white noise, but it is far from clear what
causes the pattern seen in them.  It could arise from a deficiency  of
our astrometric model.  The addition of an acceleration term to our model
largely removes the below-zero mean from \DEC\ residuals for 2004 and 2005, 
but seven of the eight \DEC\ residuals from 2001 October through 2003 September
remain positive, which suggests that this addition is not sufficient to leave
only white noise in the residuals.  Also, as noted earlier, the constant acceleration 
obtained from Solution~2 is not consistent with the \IMP\ positions derived from
the four earlier (1991-1994) VLBI sessions.  Therefore, the addition of 
the acceleration term to our model seems
unjustified.  Neither does it appear that any reasonable enhancement of our
orbit model 
(with, e.g., eccentricity, a third body, or anisotropic emission)
could remove the largest systematic features in the residuals in
Figure~\ref{fresvt}, features which each span about two years and include at 
least six epochs that are well distributed in phase.  
It is an even more open-ended task to rule out
the possibility that the systematic residuals are caused by 
inaccuracies in our data-reduction scheme that could also contribute
important systematic errors to our astrometric parameter estimates.  
What we can say is that, first, we cannot identify any such
inaccuracies that could plausibly explain our \DEC\ residuals, and, 
second, the lack 
of comparably large and systematic residuals in our fits for the 
relative positions of our three reference sources, despite their larger 
separations on the sky,\footnote
{During the above-mentioned period 2001 October through 2003 September, 
in which the \IMP\ positions have a mean residual of $\sim$+0.5~mas in 
declination, the mean offsets of the three reference sources relative to each other
are all less than  0.2~mas in any direction (as can be inferred from the 
position residuals plotted in Figure 6 of \citetalias{GPB-III}).}
argues strongly that the origin of these features lies not in modeling errors
(such as errors in our corrections for the effects of the ionosphere), 
but rather in changes
in the physical location of the \IMP\ radio emission region.  The statistical
properties and possible physical interpretations of these changes are
explored in Papers VI\nocite{GPB-VI} and VII\nocite{GPB-VII}.

Furthermore, it is not unreasonable that the motion of that radio emission about
a mean orbit fit to its positions would appear to be noise with a nonzero
autocorrelation function.  The major obstacle to our accounting for the
effects of such systematic residuals with a suitable noise model is that we lack 
adequate prior knowledge of the autocorrelation function.  It could plausibly be 
monotonic, periodic, or quasiperiodic---though the periodogram analysis of the
residuals (described above) failed to identify any significant periodicity.
Even for monotonically decreasing functions of time separation, there is
inadequate constraint on the appropriate functional form.  The inadequacy is
worsened by our having just a few effectively independent samples of any 
noise process whose autocorrelation width is one or two years.  
We therefore collaborated with Jingchen Liu and Xiao-Li Meng of Harvard's
Department of Statistics on statistical exploration and Bayesian analyses
of our position estimates.  These analyses \citep{LiuJ2008} 
revealed that allowance for the autocorrelation could increase those SSEs 
by a factor of two or so, depending, of course, on the prior probability 
distributions we adopt for the parameters describing the autocorrelation.  
Rather than discussing and interpreting those uncertain results, we
take here a simpler approach.  

Suppose we were to assume that, for the purpose of estimating \IMP's proper
motion, we can regard our observations as yielding only one independent
measurement every two years.  Given that we obtained about four positions per
year, the uncertainty of our WLS proper-motion estimate would be increased by
a factor of up to $\sqrt{8}$.  But such a ``worst-case" increase almost surely
overestimates the effect of the autocorrelation, given that there is no two-year
subset of our data during which the mean \RhA\ or \DEC\ residual was
larger than the rms difference between consecutively measured position
coordinates.  Consequently, to estimate the statistical contribution to the
total SE of our proper-motion estimate, we reduce the worst-case
factor of $\sqrt{8}$ to a factor of two.  Although this choice is clearly
somewhat arbitrary, we believe that any significantly larger factor would be
unrealistically conservative. 

We can check the appropriateness of our choice 
of a factor of two by making trial solutions in
which we remove an interval of data that might be affected by a relatively
large value of some noise-like but autocorrelated offset.  
Solution~7 in Table \ref{tres} is one such solution.  We obtained it by 
excluding the final five VLBI positions, which
are those whose residuals exhibit (in Figure~\ref{fresvt}) the most clearly
systematic and one-sided offsets from the model yielded by Solution~1.  The
resulting proper-motion estimate differs from that from
Solution~1 by 0.07~\masyr\ in \DEC\ (but less than 0.01~\masyr\ in \RA). 
Moreover, Solution~7 underestimates the mean declination of
the four unweighted 1991-1994 VLBI positions by 0.6~mas, 
suggesting that this solution is inferior to Solution~1, which underestimates 
their mean declination by only 0.2~mas.  Further, we can surmise from
inspection of Figure~\ref{fresvt} that the omission of no other group of five
consecutive positions would cause even this large a change.  Thus we consider
Solution~7 to be good evidence that doubling the SSEs obtained in Solution~1,
to 0.052 \masyr\ in \RA\ and 0.059 \masyr\ in \DEC, gives reasonably 
conservative values for SSEs with adequate allowance for the possible 
contributions of errors that are correlated between different epochs.  

We also increase our initial estimates of the SSEs in \IMP's position
coordinates at the reference epoch by the same 
somewhat arbitrary factor, since those estimates,
too, would be strongly affected by any autocorrelation spanning a year or two.
The other parameters in our 9-parameter solutions govern terms in our
astrometric model that are periodic with a period of either 1~yr or $\sim$24.65~d.
If autocorrelation of the noise in our position estimates is consistently
positive for time differences less than a year, the uncertainties in our
estimates of these parameters would be smaller, not larger, than they would be
without such autocorrelation.  Thus there is no reason to increase our SSE
estimates for these other parameters.

\subsubsection{Systematic errors}
\label{sssse}

We turn now to other contributions to our uncertainty:
propagation delays, inadequately mapped source structure, and inaccuracies
in the parameter values used in the reduction of our data, such as inaccuracies 
in Earth-orientation parameters and antenna positions.  Based on the observed
stability of the relative positions of our three reference sources, we place
upper bounds on all errors that are not intrinsic to \IMP.  (The more direct 
approach, summing in quadrature the estimated SEs due to the individual known
sources of systematic error, would be less comprehensive and hence less 
reliable.)

What, in particular, can we learn from our reference-source results?  In
\citetalias{GPB-III} we estimate the relative proper motions between the chosen reference
points in our three reference sources.   Allowing for 1-$\sigma$ uncertainties, we 
find that the ``C1" core component of \C\ moves with respect to each of the other 
two sources by less than 0.04~\masyr\ in each coordinate.  
Since the separation from \C\ of one of these sources is twice that of \IMP, and 
that of the other is five times greater, these results suggest that all the 
sky-separation-dependent error sources listed above, 
including ionospheric propagation delays,
contribute less than $\sim$0.02~\masyr\ to the SE of \IMP's proper motion.  
Moreover, as discussed in \citetalias{GPB-III}, due to the established or inferred cosmological 
distance of each of these reference sources, we expect that the true
proper motion of their radio cores is smaller still.  It would thus
require very unlikely cancellations for the above $\sim$0.02~\masyr\ bound
to be so wrong as to significantly affect our value for the total SE of \IMP's proper motion.
A similar argument can be made concerning the errors associated with
the individual radio sources, including those associated with 
the C1 core component of \C.  In \citetalias{GPB-III}, 
over the span 1998.71-2005.54, we determine upper bounds on C1's proper motion of 
0.046~\masyr\ in \RA\ and 0.056~\masyr\ in \DEC,  in 
an extragalactic reference frame closely related to the International Celestial 
Reference Frame 2 \citep[ICRF2;][]{FeyGJ2009}.  The sizes of the observed differential
proper motions among our reference sources quoted above suggest that these
upper bounds are conservative, since part of the relative proper motions is
likely due to motions of the brightness peaks of \Ba\ and \Bb.  We therefore adopt 
these last upper bounds as the total contributions to the
SE of the corresponding coordinates of our estimate of \IMP's proper motion 
due to all uncertainties not intrinsic to \IMP.

We must also allow for 
the possibility of systematic changes in the location of the radio emission
relative to the \IMP\ primary.  The radio emission from \IMP\ is fairly 
typical of that from other RS~CVn spectroscopic binaries, which, like \IMP,
usually contain a chromospherically active, cool giant primary and an inactive
secondary \citep{GuinanG1993}.  Thus there is good reason to infer that \IMP's
radio emission is closely associated, both causally and spatially, with the
primary.  Indeed, as discussed below in \S~\ref{scomp}, the orbit we determine
for the radio emission is consistent with that of the primary.  However, that
consistency does not rule out the existence, during the span of our
observations, of a significant trend in the spatial offset between the peak of
the radio emission and the center of the primary.  In particular, we
must consider the possible effect of a slow evolution of that star's magnetic field,
since such evolution might be expected if the star has a multiyear magnetic
cycle.  The strong downward trend in the flux density of \IMP's radio emission 
(see Figure~\ref{fresvt}) is, in itself, strong evidence of some kind of evolution 
of the radio emission region.  Thus we must make some allowance
for the possibility that the position of the radio emission relative to the
stellar components of the binary has a nonzero trend.  Since the rotation period of
the primary is approximately equal to the binary orbital period (as noted above
in \S~\ref{ssspfres}),
one can expect that any offset which persists for an orbit period or longer
between the center of the radio-emitting region and the center of the primary
would also corotate with the binary, and hence that the offset's equatorial
component would largely average out in a nonrotating frame.  On the other
hand, the component normal to the equatorial plane would be identical in the
rotating and inertial frames, and so could plausibly exhibit a long-term
trend as seen from Earth.  The key question is:  How large could that trend 
plausibly have been over our observing span?

Since we have no way to measure any such trend, the best we can do is to
conservatively estimate a plausible rms magnitude for it.  Three lines of
reasoning suggest that the net change, over the 
span of our observations, of the mean offset described above is likely too 
small to contribute a major source of error.
In our images of \IMP, the mean apparent extent of the 
emission region is 1.4 $\pm$ 0.4 mas (see \citetalias{GPB-VII}), and 
only at 6 of our 35 epochs did the length exceed
2~mas, with the maximum value being 3.3~mas (on 2001 January~15).
Also, within any individual session, we never detected motions of the brightness
peak exceeding 1~mas.  More importantly, using the empirical model of the 
spatial distribution of the brightness peaks relative to the IM Peg primary 
that we develop in \citetalias{GPB-VI}, we find that 2/3 of them occur at a 
distance from the center of the primary that subtends no more than 0.8~mas.  
Thus we conclude that even during individual VLBI sessions, the offset of our 
VLBI position from the primary is unlikely to exceed $\sim$1~mas.

In addition, while there was a strong (though ``noisy") downward trend in the stellar 
radio flux density over the span of our observations,
there was no clear corresponding trend in the evolution of the shape of the
emission region.  This lack of a clear trend in shape adds to our confidence that,
over the 8.5 yr span of our VLBI observations, the net motion of the mean position of 
the emission with respect to the \IMP\ primary did not vary by as much as 
$\sim$1~mas.  Consequently, it is implausible that
mean rate of relative motion over that span
exceeded $\sim$0.12~\masyr.  Even this rate of 
angular motion seems implausibly large for an rms value, for two reasons.
Given the above maximum offset, any stellar activity cycle 
with a period significantly different from about twice the length of our 
observations would result in significantly lower drift rates.
Secondly, although our lack of knowledge of the uncertainties of the
four radio position estimates we obtained 
from VLBI observations between 1991 and 1994 led us to omit 
those estimates from our chosen Solution~1, they do provide at least some 
evidence that the timescale of any systematic drift is more than $\sim$14~yr.
We thus believe that the rate corresponding to a 
shift of one stellar radius over our 8.5-year span, 0.075~\masyr, is a
sufficiently conservative estimate for the rms drift rate.  Since the projection 
on the sky of the orbit normal is along p.a.\ = 130.5\arcdeg\ 
$\pm$ 8.6\arcdeg, we allow for an rms systematic error contribution in our 
proper-motion estimate of 0.06 \masyr\ in \RA\ and 0.05 \masyr\ in \DEC.

Systematic errors could also affect our estimates for other parameters.  Our
position estimate for the center of mass of the binary at any given reference
epoch could be biased due to the type of systematic offset discussed 
directly above.
Indeed, as discussed in \citetalias{GPB-VI}, 
such a bias may well be caused by the on-going partial occultation by the primary of the 
radio emission region that we infer surrounds it.  The inclination
of the spin axis to the plane of the sky (discussed in Sections \ref{sfinres} and \ref{scomp}),
together with the observed elongation of the scatter of our VLBI residuals
(discussed in \psix), breaks the symmetry that might otherwise have led us to
believe that the offset between the radio emission and the center
of the binary system would average out over time.  Thus we must make some
estimate of the expected value and the uncertainty of the resulting error contributed 
to the position we estimate for the 
binary center of mass at the mean epoch of our VLBI observations, 2001.29.
Because we have no reliable quantitative model of this contribution, we are forced
to make somewhat arbitrary choices for its mean value and uncertainty.  
We do so in light of
the considerations stated above in regard to the mean rate of change of
the error during our span of VLBI observations.
For simplicity, we take the expected value for the bias to be zero and its
rms error to be one-half the angular radius of the primary, directed along the 
sky projection of our inferred direction for the normal to the binary orbital plane.
To calculate the rms systematic position error at any other epoch, we assume
that this error at mean epoch 2001.29 is statistically independent of any related
systematic error in our proper-motion estimate. 
Thus, to our error at 2001.29 we add in quadrature
the product of the rms drift rate estimated above and the time difference
between the other epoch and 2001.29.  In any case, the \GPB\
relativity tests have virtually no sensitivity to any possible 
milliarcsecond-scale constant bias in our estimate of the position of \IMP;
we discuss the bias only to facilitate the future use of our position estimate for 
some other purpose, such as for helping to tie the reference frame of a 
stellar position catalog to that of an extragalactic radio source catalog.

In our parallax estimate, systematic errors could occur if our position
estimates are subject to seasonally dependent errors.  
Indeed, it's plausible that inaccuracies in our models of
the troposphere and ionosphere contribute such seasonally dependent errors to our
position measurements.  Thus we need an estimate of the size of the resulting
parallax errors, or some bound on it.  We pursued three separate approaches to
this goal.

First, because VLBI observations made during more humid,
warmer months tend to suffer larger atmospheric effects, we obtained an
additional astrometric solution (not tabulated) after excluding the positions from such observations.
Since all but one of our antennas are in the Northern Hemisphere,
we excluded all observations from April through September.  Unfortunately, this
solution (and also a complementary solution based on data obtained only from
April through September) yielded a parallax SSE 3 times (and 4 times) larger
than did the solution using all the data.  Consequently, even though the
resulting changes in our parallax estimate were up to 4.4 times the SSE of the
earlier estimate, they were each less than 1.5 times the SSE of the corresponding new
estimate, and hence not truly meaningful.  

Second, we exploited the tendency of atmospheric errors to more strongly affect 
differential \DEC\ estimates than differential \RhA\ estimates, for sources observed over 
a range of hour angles spanning the central meridian, as was the case at the central sites 
of our VLBI array.  This tendency is particularly strong in the relevant case of source pairs 
separated mainly in \DEC\ \citep [see, e.g.,][]{Pradel+2006}.  A fit to our
\RhA\ data alone yielded a parallax estimate that differed from our chosen
estimate by only 0.045~mas, with an SSE only $\sim$20\% larger than that of
the fit to all the data.  This result, coupled with those in
Figure~\ref{fresvyr}, suggests that systematic error did not significantly
degrade our estimated \DEC\ coordinates more than our \RhA\
coordinates.  We thus infer that our atmospheric models likely did not significantly degrade
our parallax estimate.

Finally, we derived a third and strongest upper bound on the
plausible size of any systematic error in our parallax estimate by
using our VLBI data to estimate the relative parallax between \Ba\ and \C, and
taking advantage of the fact that their great distances from the solar system
ensure that their true parallaxes are undetectably small.  To obtain a close
analog to the systematic error in our stellar parallax estimate, we fit to the 35
relative positions we estimated for \Ba\ with respect to \C\ the same nine-parameter
model used to model the motions of \IMP.
We also gave those differential positions the same uniform SEs that we use for \IMP.  
The resulting parallax estimate for \Ba\ is $-0.032$ $\pm$ 0.074~mas.
Alternatively, if we rescale the position SEs so that the value of $\chi^2$ per degree of 
freedom of the resulting residuals is unity, the SSE of the parallax estimate
falls to 0.026~mas.  Either way, the result suggests that the systematic error in the
\Ba\ parallax estimate is smaller than the 0.074~mas SSE of our \IMP\ results.  
Moreover, to estimate the size of the systematic error in our stellar parallax,
we should take into account that the separation of \IMP\ from \C\ is fivefold smaller 
than that of \Ba\ from \C.  Since virtually all of the seasonally dependent
contributions to our measurement errors can be expected to scale with (vector)
separation on the sky, we estimate that the systematic contribution to the
SE of our stellar parallax value is, at most, one fifth of its 0.074~mas SSE.  
We thus take 0.015~mas as a plausible upper bound on the rms systematic error 
in our stellar parallax estimate. This bound implies that systematic error makes 
a negligible contribution to the total SE of the estimate.  

The above-mentioned uniformly weighted fit to our differential 
reference-source positions also provides a test for systematic errors in our
estimates of the four sinusoidal terms in our model for the \IMP\ radio 
emission.  As would be expected in the absence of significant systematic 
error, none of the four ``test" estimates differs significantly from zero; 
the largest in absolute value is only 0.05~mas (and only 1.2~times its SSE).
In light of this test and all the others mentioned previously, we believe that 
our error allowances are consistent with our data and adequate to 
estimate the true uncertainties of our results.

\section{Final Results}
\label{sfinres}

We present in Table \ref{tfinal} our final results
from Solution~1 for all nine astrometric parameters that define our
model, using each of the following epochs: 
2005.08 (2005 February 1, the approximate midpoint of the \GPB\ science data), 
2001.29 (the mean epoch of the 1997-2005 VLBI data, 
1991.25 (the epoch of the {\em Hipparcos} Catalogue),
and 2000.0.  
We present, too, an alternative parameterization of the projected orbit:
length of its semimajor axis, position angle of the major axis
(which is also that of the node in the plane of the sky), 
axial ratio, and time of conjunction 
\citep[the one nearest the time of conjunction estimated by][]{Marsden+2005}. 
The table also contains 
the SSEs from the WLS fit and our estimated total SEs.  These
latter contain allowances for the apparent autocorrelation in 
the VLBI position errors and for other possible systematic 
errors, with their variances summed under the reasonable assumption that the 
systematic errors are independent of each other.

\begin{deluxetable}{l c c c c}
\tablecaption{Final \IMP\ parameter estimates \label{tfinal}}
\tabletypesize{\scriptsize}
\tablehead{
\colhead{Parameter~~~~~~~~}  
  & \colhead{~~~~~~~~~Estimate~~~~~}  & \colhead{~~~~SSE~~~~}  
  & \colhead{Systematic Error\tablenotemark{a}}
  & \colhead{Total SE\tablenotemark{b}}  }
\startdata
\sidehead{Non-orbit parameters:}
$\alpha$ at epoch 2005.08\tablenotemark{c} (errors in mas) &  
\Ra{22}{53}{2}{258612}   & 0.12 & 0.33 & 0.40 \\
$\delta$ at epoch 2005.08\tablenotemark{c} (errors in mas) &  
\dec{16}{50}{28}{16005} & 0.13 & 0.29 & 0.39 \\
$\mua$\tablenotemark{d} (\masyr) &  $-20.833$  & 0.026 & 0.073  & 0.090  \\
$\mud$ (\masyr) &  $-27.267$  & 0.030 & 0.074  & 0.095  \\
Parallax (mas) &  $\phantom{-}
                     10.370$  & 0.074 & $<$0.015 & 0.074 \\[3pt]
\sidehead{Linear model orbit parameters:\tablenotemark{e}}
\Asa\ (mas) &  $-0.59$  & 0.10 & $\ll$0.1 & 0.10 \\
\Asd\ (mas) &  $-0.66$  & 0.11 & $\ll$0.1 & 0.11 \\
\Aca\ (mas) &  $\phantom{-}
                 0.15$  & 0.09 & $\ll$0.1 & 0.09 \\
\Acd\ (mas) &  $-0.23$  & 0.11 & $\ll$0.1 & 0.11 \\[3pt]
\sidehead{Alternative orbit parameters:\tablenotemark{f}}
Semimajor axis (mas) &  0.89  & 0.09 & $\ll$0.1 & 0.09 \\ 
Axial ratio\tablenotemark{g}  &  0.30  & 0.13 &$\ll$0.1 & 0.13 \\ 
P.A.\ of ascending node\tablenotemark{h} (deg) &  40.5  & 8.6 & $\ll$8 & 8.6 \\ 
$T_{\rm conj}$ (heliocentric JD)\tablenotemark{i} &  2450342.56
  & 0.44 & $\ll$0.4 & 0.44 \\[3pt]
\sidehead{Positions at alternative reference epochs\tablenotemark{f}(errors in mas):}
$\alpha$ at epoch 2001.29  &  
\Ra{22}{53}{2}{264124}   & 0.07 & 0.30 & 0.35 \\
$\delta$ at epoch 2001.29  &  
\dec{16}{50}{28}{26362} & 0.08 & 0.30 & 0.34 \\
$\alpha$ at epoch J2000  &  
\Ra{22}{53}{2}{265997}   & 0.07 & 0.35 & 0.38 \\
$\delta$ at epoch J2000  &  
\dec{16}{50}{28}{29883} & 0.09 & 0.36 & 0.40 \\
$\alpha$ at epoch 1991.25  &  
\Ra{22}{53}{2}{278694}   & 0.27 & 0.89 & 1.04 \\
$\delta$ at epoch 1991.25  &  
\dec{16}{50}{28}{53741} & 0.31 & 0.94 & 1.13 \\
\enddata
\tablenotetext{a}{ 
The uncertainties in the position and proper motion of the 
phase reference point C1 in \C\ are included here, and not in the SSE.
The uncertainty due to the possible offset and secular drift between the mean 
position of the stellar radio emission and the center of mass of the binary are
likewise included here.  However, the upper bounds on the systematic
errors in the orbit terms apply to the mean orbit of the radio emission,
and not to the corresponding orbital terms for the stellar binary.
}
\tablenotetext{b}{ Each ``total SE" is our estimate of the
parameter's SE, computed as the 
root-sum-square (RSS) of the SSE and our estimated 
systematic error.  For the position and proper-motion parameters, 
we first doubled the SSE before computing the RSS, to allow for 
correlated ``noise" in the VLBI positions.
}
\tablenotetext{c}{
The position given is the estimated position 
of the center of mass of the \IMP\ binary
at epoch JD 2453403.0 (2005 Feb 1, $\sim$2005.08), 
the approximate midpoint of the \GPB\ science data.
Along with the proper motion, the position is specified in the (J2000.0) 
coordinate system described in \S~\ref{ssdefpos} and 
\protect\citetalias{GPB-III}.
This nearly inertial, extragalactic, coordinate system is closely tied to the
International Celestial Reference Frame 2 \protect\citep[ICRF2;][]{FeyGJ2009}.
}
\tablenotetext{d}{ 
As elsewhere, $ \mu_{\alpha*} = \mu_{\alpha}\cos\delta$
(see Table~\ref{tres} note {\em a}).
}
\tablenotetext{e}{
In our linear model, the orbital contribution to \IMP's position at time $T$ is  
\Asa~sin~$[2\pi(T - T_{\rm conj})/P]$ + \Aca~cos~$[2\pi(T - T_{\rm conj})/P]$ 
in \RA\ and 
\Asd~sin~$[2\pi(T - T_{\rm conj})/P]$ + \Acd~cos~$[2\pi(T - T_{\rm conj})/P]$ 
in \DEC, where $P$ = 24.64877~d is the (fixed) orbital period and 
$T_{\rm conj}$ is the (fixed) time of conjunction, JD 2450342.905, adopted 
from \citet{Marsden+2005}.
}
\tablenotetext{f}{See text, \S~\ref{ssmodel}.
}
\tablenotetext{g}{The ratio of the minor axis to the major axis of the 
 sky-projected orbit.
}
\tablenotetext{h}{See \protect\citetalias{GPB-VI}
for illustration of the orbit geometry.  The
orbital motion on the sky is counterclockwise.
}
\tablenotetext{i}{Time of conjunction for the radio emitting region.  
The value shown is for the conjunction nearest the one 
with the primary in back, i.e., at its greatest distance from us,
for the optical orbit of \protect\citet{Marsden+2005}.
}
\end{deluxetable}

The correlations 
of the parameter estimates from our WLS fit are given in Table~\ref{tcorr}.
The high correlations among the estimates of
position and proper motion are a consequence of the displacement of the
reference epoch from the mean epoch of our data.  (At the mean epoch of our
VLBI data, 2001.29, the correlations are only $-0.024$ for the \DEC\ 
components and $-0.015$ for the \RhA\ components.)
In contrast, the dominant cause of
the correlations between the \RhA\ and \DEC\ components of these terms and
also between those of
the orbit terms is the correlation at each epoch of the noise-like errors 
in \RA\ and \DEC, whose value we set at $-0.314$ based on the iterative 
procedure described in \S~\ref{ssmodel}.  The relatively small 
magnitude of the correlations among  
the other estimates is the result of our successful efforts to schedule
our observations at epochs well distributed over orbit phase and season
of the year.  We do not adjust these correlations to
account for systematic errors, primarily because we have
insufficient knowledge of the correlations between the \RhA\ and \DEC\
components of those errors.

\begin{deluxetable}{l@{\hspace{50pt}}c c c c c c c c c}
\tablecaption{Correlation matrix for the \IMP\ parameter estimates \label{tcorr}}
\tabletypesize{\scriptsize}
\tablewidth{0pt}
\tablehead{
Parameter
 & \colhead{~$\alpha$} & \colhead{~$\delta$}
 & \colhead{~$\mua$} & \colhead{~$\mud$} & \colhead{~Parallax}
 & \colhead{~~\Asa} & \colhead{~~\Asd} & \colhead{~~\Aca} & \colhead{~~\Acd} }
\startdata
$\alpha$  & $\phantom{-}$1.00 \\
$\delta$  & $-$0.30 & $\phantom{-}$1.00 \\
$\mua$    & ~$\phantom{-}${\bf 0.83}\tablenotemark{a}
          & $-$0.25 & $\phantom{-}$1.00 \\
$\mud$    & $-$0.25 & ~$\phantom{-}${\bf 0.83}\tablenotemark{a}
          & $-$0.29 & $\phantom{-}$1.00 \\
Parallax  & $-$0.12 & $-$0.05 & $-$0.19 & $-$0.07 
          & $\phantom{-}$1.00 \\
\Asa\     & $\phantom{-}$0.03 & $-$0.01 & $-$0.06 & $\phantom{-}$0.01
          & $\phantom{-}$0.05 & $\phantom{-}$1.00 \\
\Asd\     & $-$0.01 & $\phantom{-}$0.03 & $\phantom{-}$0.02 & $-$0.06
          & $\phantom{-}$0.01 & $-$0.31 & $\phantom{-}$1.00 \\
\Aca\     & $\phantom{-}$0.12 & $-$0.04 & $\phantom{-}$0.15 & $-$0.05 
          & $\phantom{-}$0.02 & $\phantom{-}$0.12
          & $-$0.04 & $\phantom{-}$1.00 \\
\Acd\     & $-$0.04 & $\phantom{-}$0.12 & $-$0.06 & $\phantom{-}$0.15 
          & $\phantom{-}$0.04 & $-$0.03
          & $\phantom{-}$0.12 & $-$0.31 & $\phantom{-}$1.00 \\ 
\enddata
\tablenotetext{a}{The two largest table entries are highlighted in boldface. }
\end{deluxetable}

\section{Comparison of Results with Previous Estimates}
\label{scomp}

\begin{deluxetable}{l@{\hspace{50pt} } c@{ } c@{ } c@{ } c}
\tablecaption{Comparison with other results \label{tcomp}}
\tabletypesize{\footnotesize}
\tablewidth{0pt}
\tablehead{
Result
& Epoch
& \multicolumn{2}{c}{Proper motion} & \colhead{Parallax} \\
\colhead{ } & { }
  & \colhead{\mua}  & \colhead{\mud}  & \colhead{$\pi$} \\
  { } & { }
  & \colhead{(\masyr)}  & \colhead{(\masyr)}  & \colhead{(mas)}
  }
\startdata
This paper  & 2001.29             
 & $-$20.83 $\pm$ 0.09  & $-$27.27 $\pm$ 0.09 & ~~10.37 $\pm$ 0.07\\
\citet{Lestrade+1999}  & 1992.92             
 & $-$20.59 $\pm$ 0.46  & $-$27.53 $\pm$ 0.40  & ~~10.28 $\pm$ 0.62 \\
{\em Hipparcos} Catalogue   & 1991.25             
 & $-$20.97 $\pm$ 0.61  & $-$27.59 $\pm$ 0.57  & ~~10.33 $\pm$ 0.76 \\
\citet{vanLeeuwenF2008}  & 1991.25 
 & $-$20.73 $\pm$ 0.28  & $-$27.75 $\pm$ 0.27  & ~~11.17 $\pm$ 0.33 \\
Tycho-2 Catalogue & $\sim$1960\tablenotemark{a}\phantom{.25}
 & $-$21.4\phantom{0} $\pm$ 1.0\phantom{0}  
 & $-$26.3\phantom{0} $\pm$ 1.0\phantom{0}  & ~~ \nodata \\
\enddata
\tablenotetext{a}{ 
We assign this approximate epoch because, for our own
similar optical data set, for both \RA\ and
\DEC\ the correlation between our proper-motion estimate and 
our proper-acceleration estimate vanishes near this epoch.
Consequently, when we also estimate \IMP's proper acceleration from the
optical data, our proper-motion SSEs are smallest near this epoch.  
}
\end{deluxetable}

Our results are compared with those from several other sources in Table~\ref{tcomp}.  
Our results for proper motion and parallax are consistent with those of 
\citet{Lestrade+1999} within their (larger) SEs.  Both of these
sets of results are in agreement with the corresponding values in
the  {\em Hipparcos} Catalogue 
\citep{PerrymanESAshort1997} to within
the latter's SEs, which are slightly larger than those
of Lestrade et al. 
However, there is some disagreement with the \IMP\
results of the new {\em Hipparcos} reduction of 
\citet{vanLeeuwenF2007,vanLeeuwenF2008}.  For \IMP's  proper motion, 
the errors yielded by the new reduction are about threefold larger than ours, 
although 30\% to
50\% smaller than those of Lestrade et al.  Our results agree with those in
\citet{vanLeeuwenF2008} in \mua\ to within the combined SE,
but disagree by 1.6 times the combined SE in \mud\ and 
2.4 times the combined SE in parallax.   
The astrometric orbit of the binary, not allowed for in either of the two 
{\em Hipparcos} analyses, could be an additional
source of systematic error in the {\em Hipparcos} parallax values
(and also in the results of Lestrade et al.). 
In any case, the 0.48~\masyr\ disagreement in \mud\ is of no consequence
for the measurement of the (north-south) relativistic geodetic effect achieved by 
the \GPB\  mission, $-6601.8 \pm 18.3$ mas \citep{Everitt+2011}.  
Furthermore, the disagreement in \mud\ estimates, like that in the parallax estimates,
has no direct or significant effect on \GPB's measurement of the smaller
frame-dragging effect, which manifests itself as a purely eastward drift of
\GPB's on-board gyros with respect to \IMP; 
this result was $-37.2 \pm 7.2$ mas \citep{Everitt+2011}.

The comparison of our (VLBI) \mua\ and \mud\ results with the corresponding 
{\em Hipparcos} (optical) results also provides a check on the
size of the systematic error due to the drift of \IMP's radio emission
with respect to the center of mass of the binary system.
However, that check is not particularly useful to us, 
since the 0.5 \masyr\ (1.6-$\sigma$) disagreement in \DEC\ makes the check at 
least tenfold less precise than the bound for which we argued in \S~\ref{sserror}.

We also looked for evidence of proper acceleration in a set of optical
astrometric positions that were collected and rotated onto a common reference
frame for the purpose of computing the proper motions in the Tycho-2 Catalogue
\citep{Hog+2000}.  We performed two WLS fits to the 14 optical
\IMP\ positions given to us by N.\ Zacharias (2006, priv.\ comm.), which 
spanned 1897 to 1991.  (The final year contains the positions
from the {\em Hipparcos} and Tycho observations, which are far more accurate than
the others, most of which have 100~mas to 300~mas SEs.)    
Our four-parameter fit (for position and proper motion only) yields a proper 
motion in agreement with the Tycho-2 value to within 0.13~\masyr\ in each
coordinate, which is a small fraction of the estimated  
$\sim$2.5~\masyr\ precision of the Tycho-2 proper motions.  
(The results are not identical due,
at least in part, to differences between the 
sets of optical positions included in the two reductions.)
More importantly, when we also estimate a proper acceleration, we find it to
be consistent with zero  to within the $\sim$0.09~\masyryr\ SSE of
each coordinate of that estimate.  This result rules out the (unlikely)
possibility that, on account of such an acceleration, the 
{\em Hipparcos} proper motion might be greatly in error at the
\GPB\ epoch, nearly 14~years later than the {\em Hipparcos} mean epoch.  
Combined with the {\em Hipparcos} proper motion, the
acceleration estimate thus provides a completely independent and purely 
optical check on our main proper-motion estimate.  The uncertainty of this 
check is, however, at the level of $\sim$2~\masyr, which is 
$\sim$20-fold larger than that of our own estimate.  

The comparison of our \IMP\ position at epoch 1991.25 with the Hipparcos
position at the same epoch reveals a $\sim$2.4~mas
discrepancy, almost purely in \DEC, that is more than twice the combined SEs.  
Obtaining our result at that epoch
required an extrapolation (back in time) of nearly six years from
the start of our main data set.  Considering the possible difficulties involved
in ensuring that the two celestial reference frames are truly aligned to adequate
accuracy, and also the problems generally associated
with the extrapolation of data, 
we will not attempt to interpret the discrepancy, or even 
judge if it is truly significant.

Unlike the optical results discussed above, spectroscopic investigations of 
\IMP\ yield three of the estimated parameters of our VLBI-derived orbit.
\citet{Marsden+2005} estimate the parameters of circular orbits 
\citep[by assumption, based on the 0.006 $\pm$ 0.007 estimate of eccentricity
by][] {BerdyuginaIT1999} for both stellar components.  Combining the
spectroscopic results with estimates of the orbital inclination and parallax of 
the system, we can compute the angular sizes of these
orbits.  Berdyugina et al.\ determine the orbital inclination to lie
between 65\arcdeg\ and 80\arcdeg, while \citet{Lebach+1999} found a 
lower bound of $\sim$55\arcdeg.  In comparison, the 0.30 $\pm$ 0.13 axial 
ratio of our VLBI-derived projected orbit (Table~\ref{tfinal}), 
corresponds to an inclination of 73\arcdeg\ $\pm$ 8\arcdeg.  The good 
agreement is consistent with the radio
orbit having the same inclination as the optical orbit, as would be expected
under the plausible assumption that the effects of any anisotropy of the
emission process or any partial stellar occultation of the emission region do
not significantly affect the shape of our VLBI-derived orbit.  Combining our
inclination and parallax estimates with Marsden et al.'s $a$~sin~$i$ estimate
for each component leads to projected orbit semimajor axes of 
0.84 $\pm$ 0.03~mas for the primary and 
1.53 $\pm$ 0.06~mas for the secondary.  Thus the 0.89 $\pm$ 0.09~mas semimajor
axis of our radio orbit agrees with that of the primary, but differs significantly
(by 6 times the combined SE) from that of
the secondary.  This result is consistent with our expectation that the 
radio-emitting region is more closely associated with \IMP's primary than with
its secondary.  This result is also qualitatively
analogous to the finding of \citet{Lestrade+1993} for the Algol system:
The motion of its radio emission on the sky is consistent with the
optically-determined orbital parameters of the active evolved star of the close binary.  We
can also compare the time of conjunction implied by our orbit with the
corresponding time found by Marsden et al.  Our estimate is 0.34 $\pm$ 0.44~d
earlier.  Thus our orbit for the radio emission is not only the same size as
that of the primary, but also in the same phase, to within our SE.
In light of our estimates of parallax and orbit size, we can set a 0.78~d 
1-$\sigma$ upper bound on the phase difference, which corresponds to a physical 
distance of 3.3~\Rsol\ between the center of the primary and the mean 
position of the radio emission.  The radius of the primary, as estimated by
\citet{BerdyuginaIT1999}, is 13.3 $\pm$ 0.6 ~\Rsol.  Consequently, 
the center of the radio emission is on average offset 
in phase from the estimated center of
the primary by no more than about one-fourth of the latter's radius.

\section{Conclusions}
\label{sconc}

1.  From our series of 35 VLBI sessions spanning 1997 January to 2005 July,
we obtained weighted least-squares estimates for the position, proper motion,
parallax, and sky-projected circular orbit of the radio emission from \IMP\
(see Table~\ref{tfinal}).

2.  The accuracy of these parameter estimates is limited primarily by the noise-like scatter 
in our VLBI position measurements for \IMP.  This scatter is not caused primarily by
measurement error, but rather by apparently random offsets of the 
brightness peaks of the stellar emission from any Keplerian orbit that can be fit to these 
peaks. 

3.  For \IMP's proper-motion parameters, and for those specifying its center-of-mass 
position at epoch, we allow for increased statistical error due to the apparent 
correlations in the position residuals for sessions separated by less than 1~yr.  
We also allow for a possible nonzero mean offset and systematic drift of the stellar 
radio emission with respect to the center of mass of the binary.

4.  Our parameter estimates for \IMP's proper motion and parallax
are each consistent with previous optical and VLBI estimates
to within the appropriate combined standard errors, with two
exceptions:  (i) The \DEC\ component of our proper motion and
that of \citet{vanLeeuwenF2008} disagrees by 0.5 $\pm$ 0.3 \masyr,  and
(ii) Our parallax estimate, while consistent with that of \citet{Lestrade+1999}
and that in the {\em Hipparcos} Catalogue, disagrees by 0.80 $\pm$ 0.34 ~mas
with the revised  {\em Hipparcos} result of \citet{vanLeeuwenF2008}.

5.  The size and phase of the orbit we fit to the stellar radio emission is consistent 
with that determined for the \IMP\ primary from optical spectroscopy by
\citet{Marsden+2005}.  

6.  Our parameter estimates are sufficiently accurate to ensure that 
the uncertainty in \IMP's proper motion makes only a very small contribution to 
the uncertainty of the \GPB\ relativity tests. 

\acknowledgements 
ACKNOWLEDGMENTS.  
We thank Bob Campbell for the original version of our WLS fitting
software and for his help in making modifications needed for our work.  
We thank John Chandler, too, for helping with this task and for providing the
planetary ephemeris used in our parallax calculations.
We thank Jingchen Liu and Xiao-Li Meng for Bayesian statistical analyses of 
our VLBI position estimates.  We also
thank Norbert Zacharias for discussions and for providing optical astrometric
data in the reference frame of the Tycho-2 Catalogue. 
Finally, we thank the referee for a careful reading of the paper and for 
valuable comments.

This research was
primarily supported by NASA, through a contract from Stanford University
to the Smithsonian Astrophysical Observatory (SAO), 
as well as a major subcontract from SAO
to York University, and lesser subcontracts from SAO to the Swiss Federal
Institute of Technology Zurich (ETH-Z), Tennessee State University,
the University of Pittsburgh, and the U.S. Naval Observatory.
The National Radio Astronomy Observatory
(NRAO) is a facility of the National Science Foundation operated under
cooperative agreement by Associated Universities, Inc.
The DSN is operated by JPL/Caltech,
under contract with NASA.  We have made use of NASA's Astrophysics Data
System Abstract Service, initiated, developed, and maintained at SAO.


\begin{thebibliography}{26}
\expandafter\ifx\csname natexlab\endcsname\relax\def\natexlab#1{#1}\fi

\bibitem[{{Bartel} {et~al.}(2012){Bartel}, {Bietenholz}, {Lebach}, {Lederman},
  {Petrov}, {Ransom}, {Ratner}, \& {Shapiro}}]{GPB-III}
{Bartel}, N., {Bietenholz}, M.~F., {Lebach}, D.~E., {Lederman}, J.~I.,
  {Petrov}, L., {Ransom}, R.~R., {Ratner}, M.~I., \& {Shapiro}, I.~I. 2012,
  this issue (Paper III)

\bibitem[{{Berdyugina} {et~al.}(2000){Berdyugina}, {Berdyugin}, {Ilyin}, \&
  {Tuominen}}]{Berdyugina+2000}
{Berdyugina}, S.~V., {Berdyugin}, A.~V., {Ilyin}, I., \& {Tuominen}, I. 2000,
  \aap, 360, 272

\bibitem[{{Berdyugina} {et~al.}(1999){Berdyugina}, {Ilyin}, \&
  {Tuominen}}]{BerdyuginaIT1999}
{Berdyugina}, S.~V., {Ilyin}, I., \& {Tuominen}, I. 1999, \aap, 347, 932

\bibitem[{{Bietenholz} {et~al.}(2012){Bietenholz}, {Bartel}, {Lebach},
  {Ransom}, {Ratner}, \& {Shapiro}}]{GPB-VII}
{Bietenholz}, M.~F., {Bartel}, N., {Lebach}, D.~E., {Ransom}, R.~R., {Ratner},
  M.~I., \& {Shapiro}, I.~I. 2012, this issue (Paper VII)

\bibitem[{{ESA}(1997)}]{PerrymanESAshort1997}
{ESA}. 1997, {The HIPPARCOS and TYCHO catalogues,
  SP-1200} (Noordwijk, Netherlands: ESA)

\bibitem[{{Everitt} {et~al.}(2011){Everitt}, {Debra}, {Parkinson}, {Turneaure},
  {Conklin}, {Heifetz}, {Keiser}, {Silbergleit}, {Holmes}, {Kolodziejczak},
  {Al-Meshari}, {Mester}, {Muhlfelder}, {Solomonik}, {Stahl}, {Worden},
  {Bencze}, {Buchman}, {Clarke}, {Al-Jadaan}, {Al-Jibreen}, {Li}, {Lipa},
  {Lockhart}, {Al-Suwaidan}, {Taber}, \& {Wang}}]{Everitt+2011}
{Everitt}, C.~W.~F., {et~al.} 2011, \prl, 106, 221101

\bibitem[{{Fey} {et~al.}(2009){Fey}, {Gordon}, \& {Jacobs}}]{FeyGJ2009}
{Fey}, A.~L., {Gordon}, D., \& {Jacobs}, C.~S., (eds.) 2009, {IERS Technical Note
  No.~35}, http://www.iers.org/IERS/EN/Publications/TechnicalNotes/tn35.html
  (Frankfurt: Verlag des Bundesamts f\"ur Kartographie und Geod\"asie)

\bibitem[{{Franciosini} \& {Chiuderi Drago}(1995)}]{FranciosiniC1995}
{Franciosini}, E., \& {Chiuderi Drago}, F. 1995, \aap, 297, 535

\bibitem[{{Guinan} \& {Gim{\'e}nez}(1993)}]{GuinanG1993}
{Guinan}, E.~F., \& {Gim{\'e}nez}, A. 1993, in Astrophysics and Space Science
  Library, Vol. 177, Astrophysics and Space Science Library, ed. J.~{Sahade},
  G.~E. {McCluskey}, Jr., \& Y.~{Kondo}, 51--110

\bibitem[{{H{\o}g} {et~al.}(2000){H{\o}g}, {Fabricius}, {Makarov}, {Urban},
  {Corbin}, {Wycoff}, {Bastian}, {Schwekendiek}, \& {Wicenec}}]{Hog+2000}
{H{\o}g}, E., {et~al.} 2000, \aap, 355, L27

\bibitem[{{Lebach} {et~al.}(1999){Lebach}, {Ratner}, {Shapiro}, {Ransom},
  {Bietenholz}, {Bartel}, \& {Lestrade}}]{Lebach+1999}
{Lebach}, D.~E., {Ratner}, M.~I., {Shapiro}, I.~I., {Ransom}, R.~R.,
  {Bietenholz}, M.~F., {Bartel}, N., \& {Lestrade}, J.-F. 1999, \apjl, 517, L43

\bibitem[{{Lebach} {et~al.}(2012){Lebach}, {Bartel}, {Bietenholz}, {Campbell},
  {Gordon}, {Lederman}, {Lestrade}, {Ransom}, {Ratner}, \& {Shapiro}}]{GPB-IV}
{Lebach}, D.~E., {et~al.} 2012, this issue (Paper IV)

\bibitem[{{Lestrade} {et~al.}(1993){Lestrade}, {Phillips}, {Hodges}, \&
  {Preston}}]{Lestrade+1993}
{Lestrade}, J.-F., {Phillips}, R.~B., {Hodges}, M.~W., \& {Preston}, R.~A.
  1993, \apj, 410, 808

\bibitem[{{Lestrade} {et~al.}(1999){Lestrade}, {Preston}, {Jones}, {Phillips},
  {Rogers}, {Titus}, {Rioja}, \& {Gabuzda}}]{Lestrade+1999}
{Lestrade}, J.-F., {Preston}, R.~A., {Jones}, D.~L., {Phillips}, R.~B.,
  {Rogers}, A.~E.~E., {Titus}, M.~A., {Rioja}, M.~J., \& {Gabuzda}, D.~C. 1999,
  \aap, 344, 1014

\bibitem[{{Lestrade} {et~al.}(1995){Lestrade}, {Jones}, {Preston}, {Phillips},
  {Titus}, {Kovalevsky}, {Lindegren}, {Hering}, {Froeschle}, {Falin},
  {Mignard}, {Jacobs}, {Sovers}, {Eubanks}, \& {Gabuzda}}]{Lestrade+1995}
{Lestrade}, J.-F., {et~al.} 1995, \aap, 304, 182

\bibitem[{{Liu}(2008)}]{LiuJ2008}
{Liu}, J. 2008, PhD thesis, Harvard University, Cambridge

\bibitem[{{Marsden} {et~al.}(2005){Marsden}, {Berdyugina}, {Donati}, {Eaton},
  {Williamson}, {Ilyin}, {Fischer}, {Mu{\~n}oz}, {Isaacson}, {Ratner}, {Semel},
  {Petit}, \& {Carter}}]{Marsden+2005}
{Marsden}, S.~C., {et~al.} 2005, \apjl, 634, L173

\bibitem[{{Mayor} \& {Mazeh}(1987)}]{MayorM1987}
{Mayor}, M., \& {Mazeh}, T. 1987, \aap, 171, 157

\bibitem[{{Pradel} {et~al.}(2006){Pradel}, {Charlot}, \&
  {Lestrade}}]{Pradel+2006}
{Pradel}, N., {Charlot}, P., \& {Lestrade}, J.-F. 2006, \aap, 452, 1099

\bibitem[{{Press} {et~al.}(1992){Press}, {Teukolsky}, {Vetterling}, \&
  {Flannery}}]{Press+1992}
{Press}, W.~H., {Teukolsky}, S.~A., {Vetterling}, W.~T., \& {Flannery}, B.~P.
  1992, {Numerical recipes in FORTRAN. The art of scientific computing,
2nd edition}
  (Cambridge: University Press)

\bibitem[{{Ransom} {et~al.}(2012{\natexlab{a}}){Ransom}, {Bartel},
  {Bietenholz}, {Lebach}, {Lederman}, {Luca}, {Ratner}, \& {Shapiro}}]{GPB-II}
{Ransom}, R.~R., {Bartel}, N., {Bietenholz}, M.~F., {Lebach}, D.~E.,
  {Lederman}, J.~I., {Luca}, P., {Ratner}, M.~I., \& {Shapiro}, I.~I.
  2012{\natexlab{a}}, this issue (Paper II)

\bibitem[{{Ransom} {et~al.}(2012{\natexlab{b}}){Ransom}, {Bartel},
  {Bietenholz}, {Lebach}, {Lestrade}, {Ratner}, \& {Shapiro}}]{GPB-VI}
{Ransom}, R.~R., {Bartel}, N., {Bietenholz}, M.~F., {Lebach}, D.~E.,
  {Lestrade}, J.-F., {Ratner}, M.~I., \& {Shapiro}, I.~I. 2012{\natexlab{b}},
  this issue (Paper VI)

\bibitem[{{Shapiro} {et~al.}(2012){Shapiro}, {Bartel}, {Bietenholz}, {Lebach},
  {Lestrade}, {Ransom}, \& {Ratner}}]{GPB-I}
{Shapiro}, I.~I., {Bartel}, N., {Bietenholz}, M.~F., {Lebach}, D.~E.,
  {Lestrade}, J.-F., {Ransom}, R.~R., \& {Ratner}, M.~I. 2012, this issue,
  (Paper I)

\bibitem[{{Tokovinin} {et~al.}(2006){Tokovinin}, {Thomas}, {Sterzik}, \&
  {Udry}}]{Tokovinin+2006}
{Tokovinin}, A., {Thomas}, S., {Sterzik}, M., \& {Udry}, S. 2006, \aap, 450,
  681

\bibitem[{{van Leeuwen}(2007)}]{vanLeeuwenF2007}
{van Leeuwen}, F. 2007, \aap, 474, 653

\bibitem[{{van Leeuwen}(2008)}]{vanLeeuwenF2008}
---. 2008, VizieR Online Data Catalog, 1311, 0

\end{thebibliography}
\end{document}